\documentclass[aip, cha,reprint,amsmath,amsfonts]{revtex4-1}

\usepackage{amssymb}
\usepackage{amsmath}
\usepackage{bm}
\usepackage{paralist}
\usepackage{graphicx}
\usepackage{enumerate}
\usepackage{tikz}
\usepackage{graphicx}
\usepackage{verbatim}
\usepackage{etoolbox}
\usepackage{nicematrix}

\usetikzlibrary{matrix,positioning,fit}

\usepackage{hyperref}
\hypersetup{
	colorlinks,
	citecolor=blue,
	filecolor=blue,
	linkcolor=blue,
	urlcolor=blue,
	pdfproducer={}
}

\let\bbordermatrix\bordermatrix
\patchcmd{\bbordermatrix}{8.75}{4.75}{}{}
\patchcmd{\bbordermatrix}{\left(}{\left[}{}{}
\patchcmd{\bbordermatrix}{\right)}{\right]}{}{}

\newlength{\Mylen}

\usepackage{ragged2e}
\usepackage{caption}
\usepackage{ragged2e}
\DeclareCaptionJustification{justified}{\justifying}
\usepackage{soul}	

\usepackage [autostyle, english = american]{csquotes}
\MakeOuterQuote{"}

\usepackage[normalem]{ulem}

\usepackage[percent]{overpic}

\usepackage{xcolor}

\usepackage{cprotect}

\begin{document} 
\title{Modeling active nematics via the nematic locking principle}

\author{Kevin A.~Mitchell}
\email{kmitchell@ucmerced.edu}
 \affiliation{Physics Department, University of California, Merced, CA
   95344, USA}

\author{Md Mainul Hasan Sabbir}
 \affiliation{Physics Department, University of California, Merced, CA
   95344, USA}

\author{Sean Ricarte}
 \affiliation{Physics Department, University of California, Merced, CA
   95344, USA}

\author{Brandon Klein}
\affiliation{Department of Physics and Astronomy, Johns Hopkins University, Baltimore, MD 21218, USA}

\author{Daniel A. Beller}
\affiliation{Department of Physics and Astronomy, Johns Hopkins University, Baltimore, MD 21218, USA}

\date{\today}
	
\begin{abstract}
Active nematic systems consist of rod-like internally driven subunits that interact with one another to form large-scale coherent flows.  They are important examples of far-from-equilibrium fluids, which exhibit a wealth of nonlinear behavior.  This includes active turbulence, in which topological defects in the nematic order braid around one another in a chaotic fashion.  One of the most studied examples of active nematics consists of a dense two-dimensional layer of microtubules, crosslinked by kinesin molecular motors that inject extensile deformations into the fluid.  Though numerous theoretical studies have modeled microtubule-based active nematics, no consensus has emerged on how to fully and quantitatively capture the features of the experimental system.  To better understand the theoretical foundations for modeling this system, we propose a fundamental principle we call the \emph{nematic locking principle}---individual microtubules cannot rotate without all neighboring microtubules also rotating.  Physically, this is justified by the high density of the microtubules, their elongated nature, and their corresponding steric interactions.  We assert that the nematic locking principle holds throughout the majority of the material, but breaks down in the neighborhood of topological defects and other regions of low density.  We derive the most general nematic transport equation consistent with this principle and also derive the most general term that violates it, introducing fracturing into the material.  We then examine the standard Beris-Edwards approach, commonly used to model this system, and show that it violates the nematic locking principle throughout the majority of the material due to fracturing.  We then propose a modification to the Beris-Edwards model that enforces nematic locking nearly everywhere.  This modification shuts off fracturing except in regions where the order parameter (a proxy for density) is reduced.  In these regions fracturing is turned on.  The resulting simulations in turn show strong nematic locking throughout the bulk of the material, with narrow bands of fracturing, consistent with experimental observation.  One additional advantage of enforcing nematic locking is that nontrivial stationary state solutions, common in Beris-Edwards simulations but not seen in experiments, are eliminated.
\end{abstract}

\maketitle

\section{Introduction}

Active matter consists of nonequilibrium collections of individual self-propelled agents that interact to form large-scale coherent motion~\cite{Marchetti13,Das20}.  They are physically interesting because they exhibit a wealth of nonlinear phenomena not seen in systems at thermodynamic equilibrium.  They are also technologically important due to their potential as advanced dynamic and functional materials.  Examples of active matter include large-scale systems, such as flocks of birds~\cite{Toner95}, schools of fish~\cite{Katz11}, swarms of insects~\cite{Buhl06}, or herds of wildebeests~\cite{Hueschen23}, as well as microscopic systems, such as swimming bacteria~\cite{Sokolov07,Wensink12,Dunkel13}, dynamic cell layers~\cite{Saw17,Kawaguchi17}, or engineered ensembles of biomolecules cross-linked by molecular motors~\cite{Ndlec97,Schaller10,Sanchez12,Henkin14,Giomi15,DeCamp15,Guillamat16,Doostmohammadi17,Guillamat17,Shendruk17,Ellis18,Lemma19}.  

\begin{figure} 
\includegraphics[width = \columnwidth]{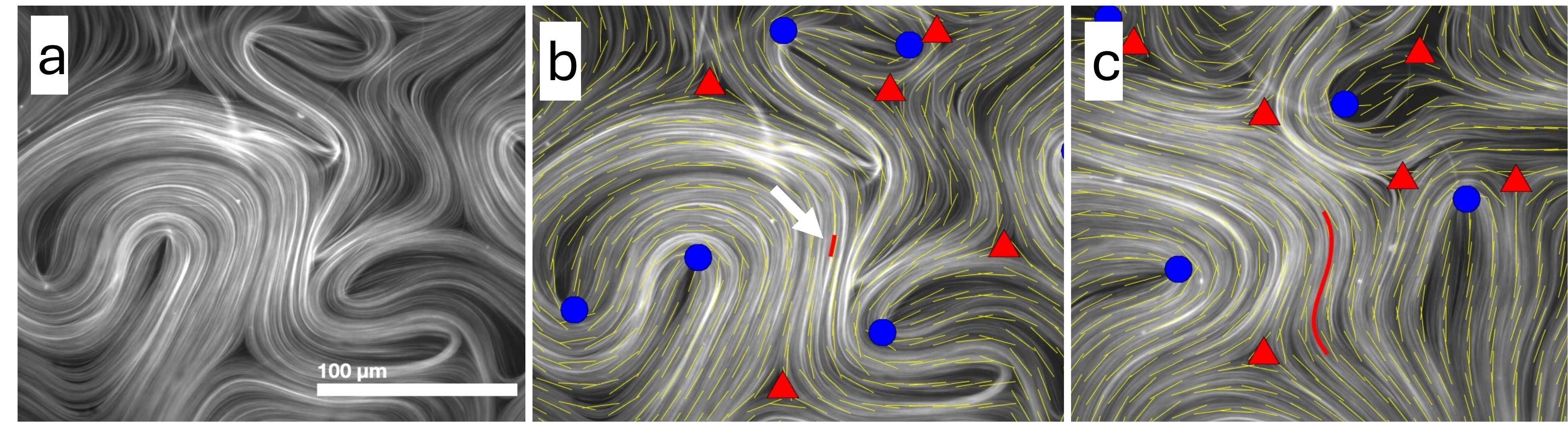}
\caption{\label{fig:striation}(a) Fluorescence image of microtubule-based active nematic, based on data from Ref.~\onlinecite{Serra23}. (b) The director field is overlayed as yellow segments. Topological defects in the director field are shown as blue circles ($+1/2$) and red triangles ($-1/2$).  A segment of a bright nematic contour is shown in red and indicated by the white arrow.  (c) The active nematic at a later time showing the advected red curve, which clearly remains a nematic contour.}
\end{figure}

This paper focuses on the modeling of microtubule-based active nematics, a prototype active material consisting of a two-dimensional (2D) layer of densely packed microtubule bundles driven by kinesin molecular motors~\cite{Sanchez12,Henkin14,DeCamp15,Serra23} (Fig.~\ref{fig:striation}a).  These motors crosslink the microtubules, driving them to slide relative one another and injecting extensile deformations into the fluid.  The process is powered by the hydrolysis of ATP.   Details of the material composition and construction can be found in, e.g.,  Refs.~\onlinecite{Henkin14,Tan19,Serra23}.  As seen in the striation pattern of Fig.~\ref{fig:striation}a, the 2D layer exhibits nematic (i.e. orientational) order due to the local alignment of the microtubule bundles.  This orientational order is described by a director field $\mathbf{n}$, with $\mathbf{n}$ physically equivalent to $-\mathbf{n}$.  As is characteristic of nematics, the nematic order breaks down at topological defects, where the director field has nontrivial winding (Fig.~\ref{fig:striation}b).  The injected ATP energy causes the system to flow in a complex manner, commonly called active turbulence, in which the topological defects braid around one another along chaotic trajectories~\cite{Tan19}.   

In continuum theories, the microtubule-based active nematic is typically described by two fields, the flow velocity $\mathbf{u}$ and the $\mathsf{Q}$-tensor, which combines the director field $\mathbf{n}$ with the scalar order parameter $S$ that measures the degree of alignment~\cite{DeGennes93,Beris94}. 
Modeling active nematics requires a partial differential equation for each field.  The flow velocity $\mathbf{u}$ is modeled by the Navier-Stokes equation with an additional activity term to inject energy into the flow.  Often, the Stokes limit of zero viscosity is taken.  The evolution of the $\mathsf{Q}$-tensor is modeled by the Beris-Edwards approach~\cite{DeGennes93,Beris94} using an equation that combines the effects of fluid motion and free energy.  The free energy includes an elastic contribution and a phase-transition term, which effect the scalar order parameter $S$ and the director orientation.  The Beris-Edwards model originated in the study of liquid crystals but is now applied to many active nematic systems.  It has been very successful in reproducing the active turbulence seen in active nematic experiments~\cite{Giomi15}.  However, it has two shortcomings when modeling microtubule-based active nematics.  First, as usually applied, the Beris-Edwards approach assumes the microtubules have uniform density.  While this may be true on average over length scales larger than the defect spacing, at smaller scales there is a wealth of fractal density variations~\cite{Mitchell21}, readily apparent in microscopy images.  Furthermore, the density at defect cores falls to zero.  The typical way of addressing this issue in practice is to view the order parameter $S$ as a rough proxy for density, at least at defect cores.  (Away from defect cores and other regions of high nematic curvature, $S$ tends to be very close to 1.)  In the present work, we retain the uniform density assumption, adopting $S$ as a proxy for density at defects and other regions of high curvature.  A second shortcoming of the standard Beris-Edwards approach is that it treats the microtubules as individual small molecules rather than forming the extended bundles they truly do.  While this is appropriate for liquid crystals and other forms of active matter, e.g. bacterial turbulence, it misses a key aspect of microtubule-based active nematics.  Namely, an extended microtubule bundle is not able to freely rotate without its entire neighborhood of bundles rotating with it, due to the strong steric interactions.  We call this principle the \emph{nematic-locking principle} because the director is locked to the flow field in a prescribed way.

In this paper, we view nematic-locking as a fundamental principle for the modeling of microtubule-based active nematics.  We derive the most general nematic transport equation for the evolution of $\mathsf{Q}$ consistent with this principle.  
Though we assume nematic locking holds throughout most of the active material, it must break down at localized regions of high curvature where the material fractures.  This fracturing is necessary for the creation of topological defects.  The associated healing process is similarly necessary for the annihilation of topological defects.  Thus, in addition to deriving the most general equation consistent with nematic locking, we derive the most general term that breaks nematic locking, i.e. gives rise to fracturing.  This analysis allows us to see which terms of the Beris-Edwards model are consistent with nematic locking and which terms break it.  The key term that requires investigation is the molecular tensor $H$, which derives from the free energy.  Part of $H$ preserves nematic locking and part of $H$ breaks nematic locking.

Via direct integration of the Beris-Edwards equations, we show that they can break nematic locking by a considerable degree throughout the majority of the material.  We then propose a modification to the Beris-Edwards equations that switches on fracturing only when $S$ is reduced from 1.  Thinking of $S$ as a proxy for density, this means that fracturing takes place when microtubule bundles become sufficiently dilute that they can separate, bend, and break.  We show that this modified equation exhibits nematic locking throughout the majority of the material domain.  Only in localized regions of high curvature, where $S$ is reduced, does it exhibit fracturing. 

Our results are consistent with two recent data-based approaches for extracting differential equations directly from experimental movies of microtubule-based active nematics~\cite{Joshi22,Golden23}.  Notably, the model equations from these studies exhibit nematic locking, supporting the idea that nematic locking takes place throughout the majority of the material.  (Neither study was sensitive to localized fracturing in the vicinity of topological defects.)

We also consider stationary states of active nematics, i.e. states in which the velocity and $\mathsf{Q}$-tensor fields do not change in time.  We argue that such active states must break nematic locking and are hence unphysical for microtubule-based active nematics.  We then show side-by-side simulations of the standard Beris-Edwards model and our modified model using the same initial condition and system parameters. The standard simulation converges to a stationary state whereas the modified version produces active turbulence. Thus, the modified Beris-Edwards model favors active turbulence over stationary states, consistent with the fact that no microtubule-based experiments have shown such stationary states.

The paper is organized as follows. Section~\ref{sec:locking} introduces the nematic locking principle.  Section~\ref{sec:NLEq} derives the general nematic transport equation for the $\mathsf{Q}$-tensor under the nematic locking principle, and also derives the general term that breaks nematic locking.  Section~\ref{sec:Expt} verifies the existence of nematic locking in experimental data.  Section~\ref{sec:BE} analyzes the Beris-Edwards model from the nematic locking perspective, showing that it can break nematic locking throughout the majority of the material.  Section~\ref{sec:BENL} introduces the central result of this work: a  modification of the Beris-Edwards model that enhances nematic locking.  In Sect.~\ref{sec:anisotropic} we investigate whether anisotropic elasticity can achieve the same effect as the modified equations, concluding that it can not.  Finally, Sect.~\ref{sec:fixed} considers stationary states, which do not evolve in time, and shows that these non-physical states are suppressed by the Beris-Edwards model with enhanced nematic locking.

\section{The locking of nematic orientation to flow}
\label{sec:locking}

A quintessential feature of the microtubule-based active nematic
system is the striation pattern formed by variations in the microtubule
density, as revealed, for example, through fluorescence microscopy
(Fig.~\ref{fig:striation}).  The striations are bundles of microtubules (rather than
individual microtubules), with an individual bundle being
an extended physical object, whose curve can be traced out in
the fluorescence image.  The striation pattern also determines the
local orientation of the nematic fluid, i.e. the director field
$\mathbf{n}$.  In short, the striation pattern reveals both extended
material objects, in the microtubule bundles, and the local director field.
These two concepts are, of course, closely related.  The director field is
everywhere tangent to the microtubule bundles, and conversely, the curve
formed by a microtubule bundle can be determined by integrating the
director field.  

 \begin{figure} 
\includegraphics[width = 1\columnwidth]{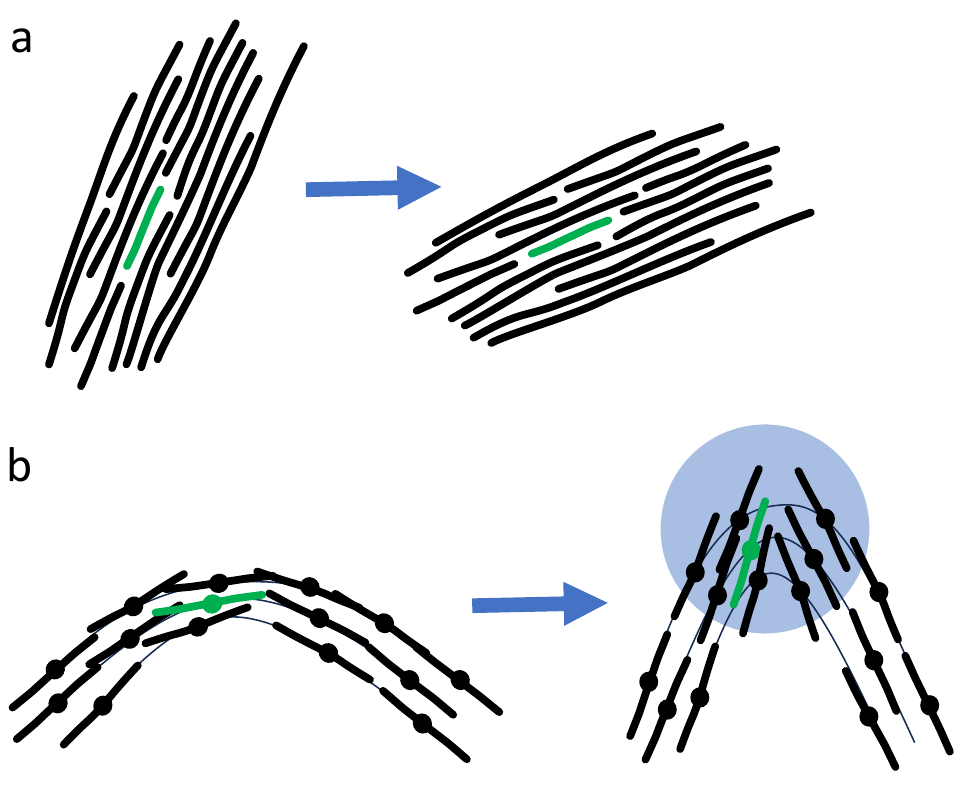}
\caption{\label{fig:bundles}. a) Cartoon of the nematic locking principle: Microtubules are packed together in extended bundles (black lines).  An individual bundle (green) cannot rotate without all neighboring bundles rotating with it due to steric interactions. b) Violating the nematic locking principle: In a region of high curvature, the density of bundles decreases so that the system fractures.  The background thin black curves are passively advected.  The centers of the bundles (dots) follow these curves, but the orientations of the bundles (e.g. the green bundle) are no longer tangent to the advected curves.  Such fracturing, however, is localized in a small domain (blue circle).}
\end{figure}

\begin{figure} 
\includegraphics[width = \columnwidth]{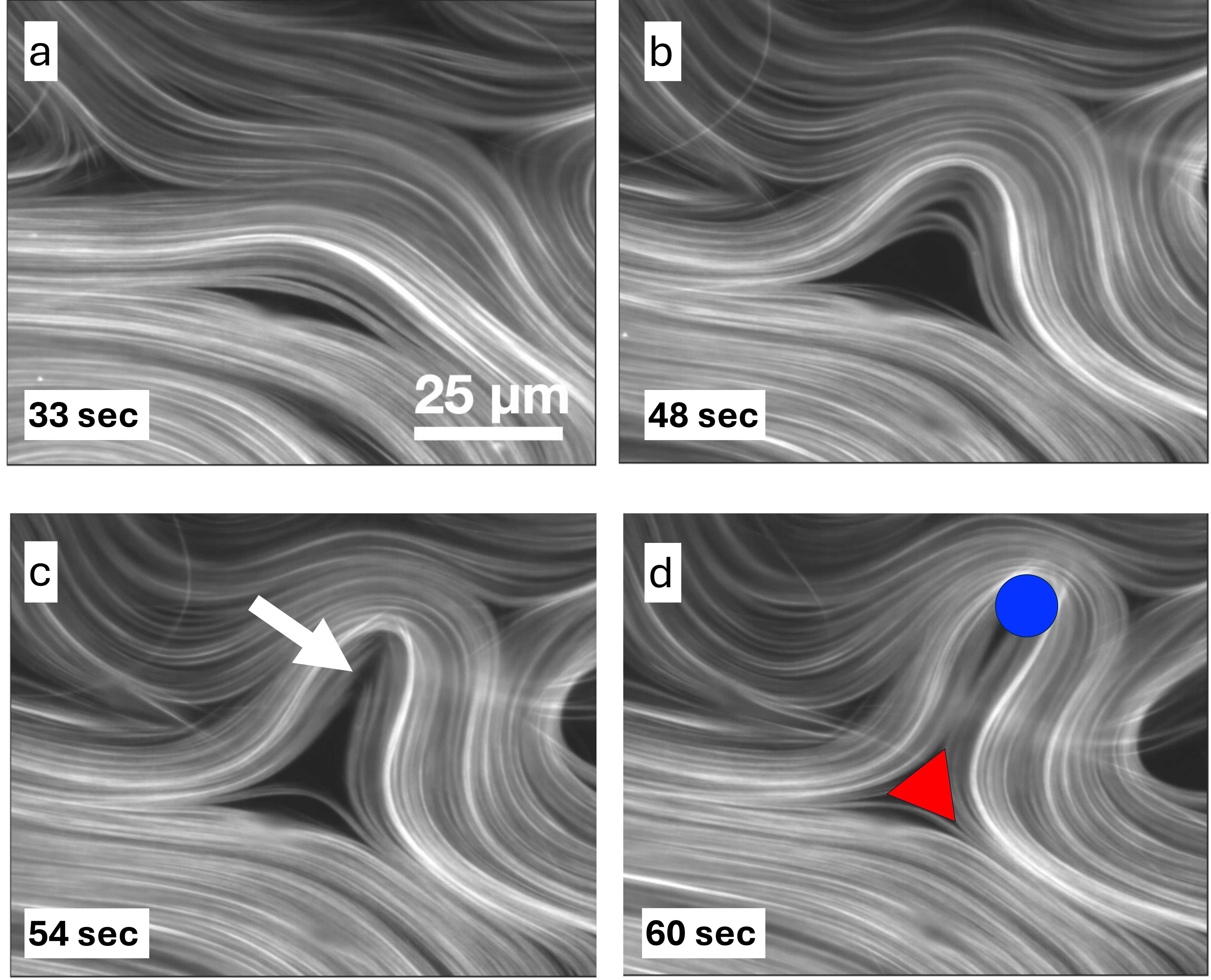}
\caption{\label{fig:fracturing} (a and b) The emergence of a high curvature region due to the folding of microtubule bundles. (c) Microtubule bundles fracture at the arrow as the curvature grows large. (d) Creation of a pair of topological defects due to the fracturing of microtubule bundles.}
\end{figure}

We refer to an integral curve of the director
field as a \emph{nematic contour}.
Consider an initial such contour determined by spatially integrating the
director field at an initial time (Fig.~\ref{fig:striation}b).  Then, advect this curve forward as
a passive material line under the fluid velocity field (Fig.~\ref{fig:striation}c).  (See Sect.~\ref{sec:Expt} for more detail on this procedure.)  At the
final time, this curve remains a nematic contour of the new director
field, as seen in Fig.~\ref{fig:striation}c.  In short, a nematic contour passively advected forward remains
a nematic contour.  Though this simple property may seem obvious, it need not
hold for a general active nematic, e.g. one with individual short nematogens that can rotate more freely.  It
crucially depends on the extended physical nature of the microtubule
bundles.  This property can be further understood in terms of steric
interactions.  The microtubules within the extended bundles are
tightly packed and hence nearly perfectly align with each other.  The microtubules within the bundle cannot rotate
independently from their neighbors in the bundle due to steric hindrance, i.e. the entire bundle must rotate or distort in order for any one microtubule to rotate (Fig.~\ref{fig:bundles}a).
Thus, the only way for the director of a parcel of material to rotate
is for the entire local neighborhood of that parcel to rotate.  This
principle may break down for shorter and/or less densely packed nematogens.

Another physical scenario that exhibits the same dynamics is when the
director field is represented by an ensemble of noninteracting thin rods
(e.g. mini toothpicks) suspended at the surface of the fluid.  Then as the fluid flows, the rods advect and rotate passively in the fluid,
i.e. the director field is passively advected by the fluid flow as
an ensemble of tiny free floating rods.  This is the relationship between the director and flow fields that we call \emph{nematic locking}, and we say that the director field
(or $\mathsf{Q}$-tensor) is \emph{locked} to the fluid flow, since knowledge of
the flow alone determines the evolution of the director field.

Nematic order locally breaks down at topological defects, which are points in the fluid with a nontrivial winding of the director field about the defect.  This winding is quantified by the topological charge, which may take on integer or half-integer values.  Microtubule systems typically only exhibit defects with topological charge $\pm 1/2$, as shown in Fig.~\ref{fig:striation}b.  The $+1/2$ defects have a comet shape in which the director field exhibits a hairpin turn.  The $-1/2$ defects have a characteristic triangular shape. 
If all directors near a defect are passively advected according to the nematic locking principle, then the defect itself moves as if it were a passively advected particle, with the same velocity as the surrounding fluid.  Thus, a perfectly nematically locked active nematic cannot exhibit defect annihilation, because no two defects can ever collide.  The same is true for defect creation as seen by running time backwards.  Thus, locking
of the director field to the fluid flow must break down in the
localized vicinity of defect creation and annihilation.  Physically,
defect creation corresponds to fracturing of the microtubule bundles
when the bundle curvature grows too large (Figs.~\ref{fig:bundles}b and \ref{fig:fracturing}).  Note that this fracturing is typically localized to a small area
of the material, where defects are created at regions of high
curvature.   More generally, nematic locking may break down any time the nematic curvature grows too large and fracturing takes place, even if a new pair is not created.  

In short, we assert---and shall provide quantitative evidence---that microtubule-based active nematics are dominated by nematic locking throughout most of the fluid, except at localized regions of high curvature where fracturing takes place that breaks nematic locking, e.g. at points of defect-pair creation and annihilation.

\section{The general nematic transport equation under the nematic
  locking assumption}
\label{sec:NLEq}

In this section we derive the general form of the nematic transport equation that respects nematic locking and also derive the general form of terms that break nematic locking.  First, recall that the dynamics of an active nematic are governed by two fields, the
fluid velocity $\mathbf{u}$ and the tensor $\mathsf{Q}$, which
describes the nematic orientation and degree of alignment according to
\begin{equation}
  \mathsf{Q} = S  (\mathbf{n} \otimes \mathbf{n} - \mathsf{I}/2), \label{r10}
\end{equation}
where $\mathbf{n}$ is the unit director pointing parallel to the
filaments and where $S \le 1$ is the scalar order parameter.  The value $S = 1$ corresponds to perfect alignment.  Due to steric
constraints, $S = 1$ nearly everywhere except at regions of high
curvature where there is local fracturing and small scale disorder in the nematogens.  This includes at topological defects where $\mathbf{n}$ is not
continuously defined and $S$ must fall to $0$.    
We also
assume incompressibility of the fluid
\begin{equation}
\nabla \cdot \mathbf{u} = 0.
\end{equation}

We present a self-contained derivation of the nematic transport
equation under the assumption of nematic locking.  Consider a small
thin rod with orientation $\mathbf{n}$ that is passively advected in a
fluid flow $\mathbf{u}$.  Since each end of the rod may be moved in
different directions by the flow,
$D \mathbf{n} / Dt = \mathbf{n} \cdot \nabla \mathbf{u}$, where
$D/Dt = \partial/\partial t + \mathbf{u} \cdot \nabla$ is the
advective derivative.  Note that we do not yet impose the constraint
$|\mathbf{n}|^2 = 1$, so that the rod is imagined to be infinitely stretchable.  Applying this result to the tensor
$\mathsf{T} = \mathbf{n} \otimes\mathbf{n}$ yields
\begin{align}
&  \frac{D}{Dt} \mathsf{T} 
                            =\left(\frac{D}{Dt} \mathbf{n} \right)
                            \otimes\mathbf{n} + \text{Tr.}  
 =  (\mathbf{n} \cdot \nabla \mathbf{u} )
    \otimes\mathbf{n} + \text{Tr.} \\
  & =  (\nabla \mathbf{u})^T \mathbf{n} \otimes \mathbf{n}
    +\text{Tr.}  =  (\mathsf{E} - \mathsf{\Omega}) \mathsf{T}
       +\text{Tr.}  \\
 & =  \mathsf{ET} + \mathsf{TE} + [ \mathsf{T}, \mathsf{\Omega}].
   \label{r12}
\end{align}
Here ``Tr.'' is the transpose of the preceding part of an expression,
$\mathsf{E} = (\nabla \mathbf{u} + \nabla \mathbf{u}^T)/2$ is the
symmetric rate-of-strain tensor, and
$\Omega = (\nabla \mathbf{u} - \nabla \mathbf{u}^T)/2$ is the
antisymmetric vorticity tensor.  Eq.~(\ref{r12}) is valid in any
dimension.  For a 2D flow, the following identity holds
\begin{equation}
\chi = \mathsf{ET} + \mathsf{TE} - \text{Tr}( \mathsf{T} \mathsf{E})
\mathsf{I}  - \text{Tr}( \mathsf{T} ) \mathsf{E} = 0,
\label{r11}
\end{equation}
which may be proved as follows.  Since the flow is incompressible, the $2 \times 2$ matrix $\chi$ is symmetric and
traceless.  The space of such matrices is
two-dimensional and spanned by the matrices 
\begin{align}
    \mathsf{P} &= (\mathbf{n} \otimes \mathbf{n} - \mathbf{n}^\perp
\otimes \mathbf{n}^\perp)/2, \\
\mathsf{R} &= (\mathbf{n} \otimes \mathbf{n}^\perp + \mathbf{n}^\perp
\otimes \mathbf{n})/2 = \mathsf{JP},
\end{align}  
where $\mathbf{n}^\perp = \mathsf{J} \mathbf{n}$, with $\mathsf{J}$ the right-handed rotation by $\pi/2$.  The matrix inner product of the left-hand side of
Eq.~(\ref{r11}) with each basis matrix is easily shown to be
$\text{Tr}(\chi \mathsf{P}) = \text{Tr}(\chi \mathsf{R}) = 0$, thereby
establishing Eq.~(\ref{r11}).
Applying  Eq.~(\ref{r11}) to Eq.~(\ref{r12}) produces
\begin{equation}
   \frac{D}{Dt} \mathsf{T} =  \text{Tr}( \mathsf{T} ) \mathsf{E}  +  [
   \mathsf{T}, \mathsf{\Omega}] + \text{Tr}( \mathsf{T} \mathsf{E})
\mathsf{I}.
\end{equation}
The constraint $|\mathbf{n}|^2 = \text{Tr}( \mathsf{T})  = 1$ can
now be imposed by adding the term $-2 \text{Tr}( \mathsf{TE}) \mathsf{T}$ to the
above, yielding
\begin{equation}
   \frac{D}{Dt} \mathsf{T} =  \mathsf{E}  +  [
   \mathsf{T}, \mathsf{\Omega}] + \text{Tr}( \mathsf{T} \mathsf{E})
(\mathsf{I} -2 \mathsf{T}).
\label{r24}
\end{equation}
By taking the trace of the above equation, we confirm $D |\mathbf{n}|^2/Dt = 0$.  Equation~(\ref{r24}) can be recast in terms of $\mathsf{P} = \mathsf{T} -
\mathsf{I}/2$ as
\begin{equation}
   \frac{D}{Dt} \mathsf{P} =  \mathsf{E}  +  [
   \mathsf{P}, \mathsf{\Omega}] - 2 \text{Tr}( \mathsf{P} \mathsf{E})\mathsf{P}.
   \label{r10}
\end{equation}
Away from defects, or other regions of high curvature, $S = 1$ and $\mathsf{Q} = \mathsf{P}$, and hence
\begin{equation}
   \frac{D}{Dt} \mathsf{Q} =  \mathsf{E} + [ \mathsf{Q}, \mathsf{\Omega}]
    -   2 \text{Tr}( \mathsf{Q} \mathsf{E}) \mathsf{Q}.
\label{r1}
  \end{equation}
This is exactly the equation extracted from the data-driven protocols
in Refs.~\onlinecite{Joshi22,Golden23}.  [See in particular Eq.~(4) in Ref.~\onlinecite{Golden23}, which is expressed in terms of the director $\mathbf{n}$ rather than $\mathsf{Q}$, and Eq.~(5) in Ref.~\onlinecite{Joshi22}.  Note that the latter includes a term $K\nabla^2 \mathsf{Q}$, though it is made clear that this term does not come from the data-driven approach, but is added afterward to avoid numerical singularities.] Thus, the assumption of nematic locking (and of
perfect alignment) suffice to model nematic transport throughout most
of the material, away from fracturing events.

Since the nematic locking condition only constrains the evolution of
$\mathbf{n}$ and not $S$, any term in the nematic
transport equation that only changes $S$ preserves the nematic locking
condition.  In particular, any term proportional to $\mathsf{Q}$ added to Eq.~(\ref{r1}) would preserve nematic locking.  Are there any other terms that preserve nematic locking?  Since $D \mathsf{Q}/ Dt$ is a 
symmetric, traceless $2 \times 2$ matrix, it must be a linear combination of
$\mathsf{P}$ and $\mathsf{R}$, introduced above.  We transition to working with $\mathsf{Q} = S \mathsf{P}$ and 
\begin{equation}
    \mathsf{U} = S\mathsf{R} =  S(\mathbf{n} \otimes \mathbf{n}^\perp + \mathbf{n}^\perp
\otimes \mathbf{n})/2 = \mathsf{J Q},
\label{r19}
\end{equation}
because they are smooth fields.  It is easy to see that a term
proportional to $\mathsf{U}$ in the nematic transport equation would rotate $\mathbf{n}$ without changing $S$, thereby breaking nematic locking.
Thus, the general nematic transport equation that preserves nematic
locking can be written in either of the following two ways
\begin{align}
  \frac{D}{D t}  \mathsf{Q}
  &= S \mathsf{E} +  [\mathsf{Q}, \Omega ] + A(\mathsf{Q},\mathbf{u}) \mathsf{Q}\\
&=  S\mathsf{E} + [ \mathsf{Q}, \mathsf{\Omega}]
    -   2 \text{Tr}( \mathsf{Q} \mathsf{E}) \mathsf{Q} + B(\mathsf{Q},\mathbf{u}) \mathsf{Q},
    \label{r2}
\end{align}
where $A$ and $B$ are scalar functions of $\mathsf{Q}$ and
$\mathbf{u}$ and potentially their derivatives as well.  
Adding any term $G(\mathsf{Q},\mathbf{u}) \mathsf{U}$ to the above
violates nematic locking and leads to fracturing of the material.  Furthermore, this is the most general term that violates nematic locking.  

The nematic transport equation can be viewed as an equation for both the evolution of the orientation of $\mathbf{n}$ and the order parameter $S$.  We shall analyze the local rotation rate $\omega_n = D \theta/Dt$ of the director field (in the Lagrangian frame), where $\theta$ is the orientation angle of $\mathbf{n}$.   The rotation rate can be computed from
\begin{equation}
   \text{Tr}(\mathsf{R} D\mathsf{P}/Dt)  
    =  2 \omega_n \text{Tr}(\mathsf{R}^2) = \omega_n,    
\label{r16}
\end{equation}
where the first equality follows from $D \mathbf{n}/ Dt = \omega_n \mathbf{n}^\perp$. 
Applying Eq.~(\ref{r10}) to Eq.~(\ref{r16}), we find the advective contribution to $\omega_n$ is 
\begin{equation}
    \omega_A = \text{Tr}[(\mathsf{E + [\mathsf{P},\mathsf{\Omega}])R}] = \text{Tr}(\mathsf{ER}) + \omega/2 = \mathbf{n} \cdot \mathsf{E} \mathbf{n}^\perp + \omega/2,
    \label{r3}
\end{equation} 
where $\omega = \nabla \times \mathbf{u}$ is the scalar vorticity.  Note that this is just Jeffery's equation with flow alignment parameter 1~\cite{Jeffery22}.  Similarly, the fracturing contribution to $\omega_n$ due to a term $G\mathsf{U}$ added to Eq.~(\ref{r2}) is simply
\begin{equation}
    \omega_F = G/2.
    \label{r18}
\end{equation}  
We can then express the rotation rate of the director field as $\omega_n = \omega_A + \omega_F$.  The nematic locking criterion is equivalent to $\omega_F = 0$.

\section{Nematic locking in experimental data}
\label{sec:Expt}

 \begin{figure}
 \centering
\includegraphics[width = \columnwidth]{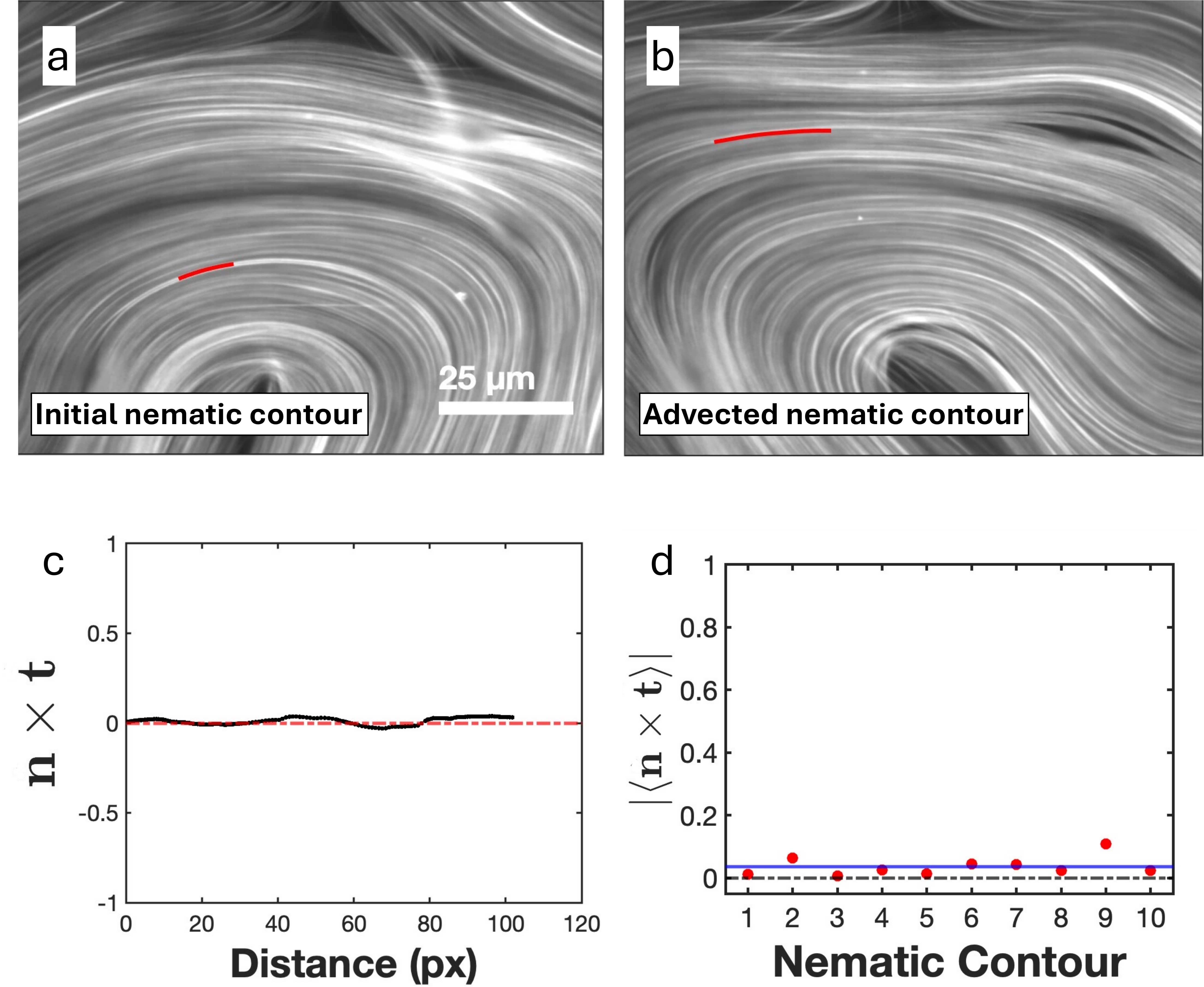}
\caption{ \label{fig:summaryExp} (a) An initial nematic contour computed from spatially integrating the director field. (b) The passively advected curve. (c) The cross product of the director field $\mathbf{n}$ and the curve tangent $\mathbf{t}$ measured along the advected curve. The fact that $\mathbf{n} \times \mathbf{t}$ along the final advected curve is approximately zero means the advected curve remains a nematic contour. (d) Summary of nematic contour analysis using experimental data for $10$ initial nematic contours. The red circles are the absolute averages of $\mathbf{n} \times \mathbf{t}$ along the length of each advected curve. The blue line is the average of $|\langle \mathbf{n} \times \mathbf{t} \rangle|$ over all 10 passively advected curves.}
\end{figure}

We show here that nematic locking can be observed and measured in experimental movies of microtubule-based active nematics.  We do this by advecting a nematic contour forward and showing that it remains (approximately) a nematic contour.  This requires independent measurements of the director and velocity fields.  We obtain this data from the experiments in Ref.~\onlinecite{Serra23}, which used two-color fluorescence microscopy.  The microtubule sample contained a high concentration of fully labeled microtubules in one color and a low concentration of sparsely labeled microtubules in another color; the latter are dilute enough that individual labeled microtubules can be distinguished. We use experimental images of fully-labeled microtubules to obtain the director field using an FFT-based (fast Fourier transform) Matlab code. We use experimental images of sparsely labeled microtubules to obtain the velocity field, using particle image velocimetry (specifically, PIVlab following the settings from  Ref.~\onlinecite{phuTran}). We chose an initial nematic contour for the analysis by spatially integrating the director field from a reference point.  We then passively advected this contour forward in time using the velocity field, with no further reference to the director field.  We chose the advection time so that the curve length grew by a factor of two.  We then checked whether the final advected curve was still a nematic contour by taking the cross product of the director field $\mathbf{n}$ and the unit tangent $\mathbf{t}$ along the advected curve. This parameter, $\mathbf{n} \times \mathbf{t}$, measures the local deviation of a true nematic contour from the passively advected curve; the cross-product is zero when the tangent and director align perfectly. In that case, the advected nematic contour remains a nematic contour.

 Figure~\ref{fig:summaryExp}a shows an initial nematic contour, and Fig.~\ref{fig:summaryExp}b shows its passive evolution once its length is doubled.  The cross product, $\mathbf{n} \times \mathbf{t}$, computed along the advected curve is also shown in Fig.~\ref{fig:summaryExp}c.  It remains very close to zero.   This process was repeated for a total of 10 curves.  (See supplemental Fig.~S1.)  The initial curves were chosen to avoid the immediate vicinity of topological defects. Figure~\ref{fig:summaryExp}d shows $\mathbf{n} \times \mathbf{t}$ averaged over the length of the advected curve for all 10 curves. The average of all $|\langle \mathbf{n} \times \mathbf{t} \rangle|$ values is $0.036$, with a standard deviation of $0.031$. Given the considerable uncertainty in accurately extracting the director and velocity fields, we consider this result essentially consistent with $0$.  This analysis reveals key evidence that the microtubule-based active nematics obey the nematic locking principle away from fracture regions.

\begin{figure}
    \centering
  \includegraphics[width= \columnwidth]{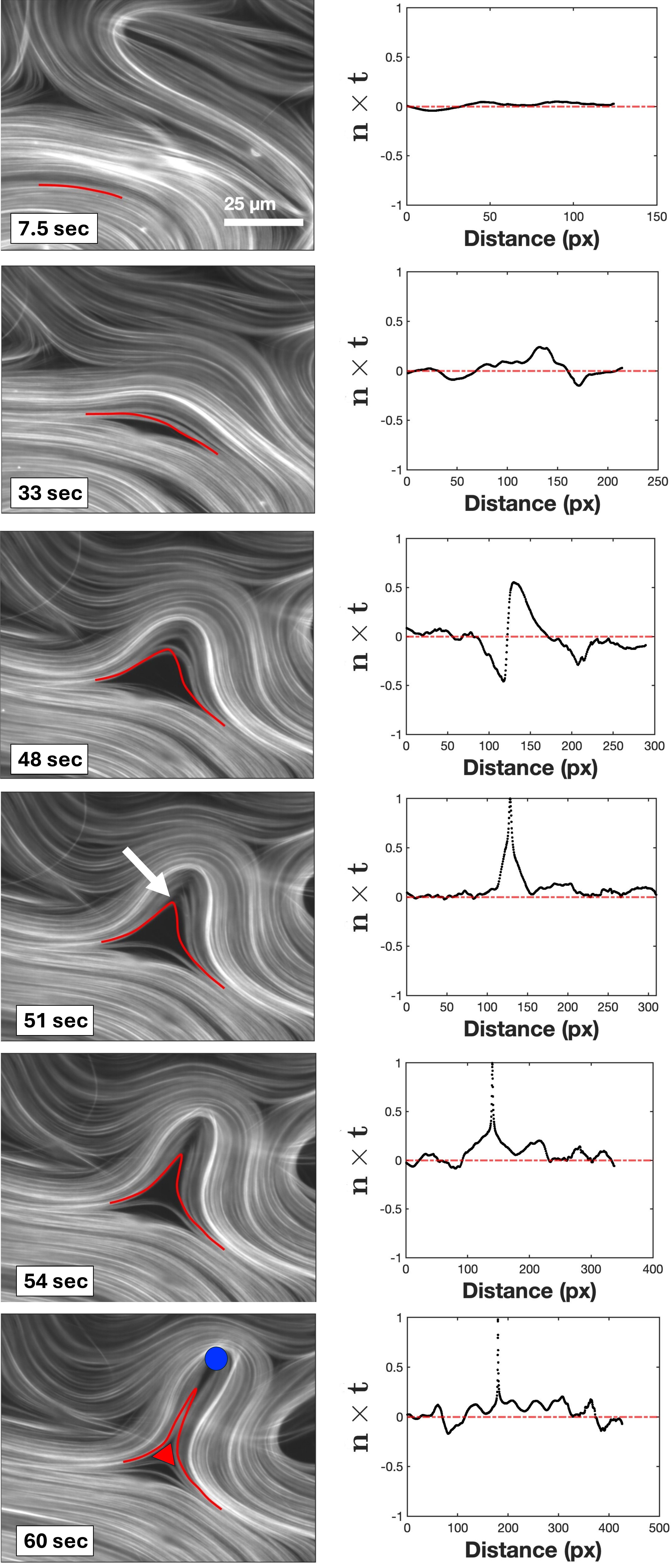}
  \centering
  \caption{ \label{fig:FracSnaps} Six images showing the fracturing of microtubule bundles that leads to the creation of a pair of topological defects. The red curve represents a passively advected curve from an initial nematic contour that enters the fracture region where a pair of topological defects is formed. The right plots show $\mathbf{n} \times \mathbf{t}$ along the advected curve at each time step. The spike in $\mathbf{n} \times \mathbf{t}$ that develops at $t = $51s represents fracturing at the sharp bend of the advected curve (white arrow).}
\end{figure}

The nematic locking principle breaks down near regions of fracturing, e.g. where topological defects are formed.  To demonstrate this, we passively advect an initial nematic contour so that it enters the fracture zone where a pair of topological defects is created.  Figure \ref{fig:FracSnaps} shows the corresponding snapshots along with the values of $\mathbf{n} \times \mathbf{t}$ at each time step.  Note the singularity in $\mathbf{n} \times \mathbf{t}$ that develops at the sharp bend in the advected curve, which is directly in the fracture zone.  Note also that the two defects end up on opposite sides of the advected curve, which would be impossible if the nematic locking principle strictly held.

\section{Analysis of a common model of active nematics}
\label{sec:BE}

\subsection{The Beris-Edwards (BE) model}

A common mathematical approach used for the evolution of $\mathsf{Q}$ is the
Beris-Edwards model, which in 2D is 
\begin{equation}
  \frac{D}{D t}  \mathsf{Q}
 = \lambda S \mathsf{E} +  [\mathsf{Q}, \Omega ] -
2 \text{Tr} (\mathsf{Q} \mathsf{E} ) \mathsf{Q}  +
\frac{1}{\gamma} \mathsf{H},
\label{r8}
\end{equation}
where $\lambda $ is the \emph{flow alignment parameter}, $\mathsf{H}$ is the
\emph{molecular tensor}, and $\gamma$ is the \emph{rotational viscosity}.  The first
three terms are in exact agreement with Eq.~(\ref{r2}), taking
$\lambda = 1$ and $B = 0$.  Thus, in regions where $\mathsf{H}$ can be ignored
(more precisely the part of $\mathsf{H}$ proportional to
$\mathsf{U}$; see Sect.~\ref{sec:NLEq}), the Beris-Edwards model exhibits nematic locking if $\lambda = 1$.
The molecular tensor is given by the equation
\begin{equation}
    \mathsf{H} = - \left(\frac{\delta F}{\delta \mathsf{Q}}
    - \frac{1}{2}\text{Tr}\frac{\delta F}{\delta \mathsf{Q}} \right).
    \label{r21}
  \end{equation}
Here, $F$ is the Landau-de Gennes free energy
  \begin{equation}
F = F^P + F^E,
  \end{equation}
composed of the phase free energy
  \begin{align}
    F^P &= \int_D   \left[ \frac{A}{2} \text{Tr}(\mathsf{Q}^2)  +
    \frac{C}{4}  \text{Tr}(\mathsf{Q}^2)^2 \right] \; da \\
    &=\int_D   \left[ \frac{A}{4} S^2  +
    \frac{C}{16}  S^4 \right] \; da, \label{r17}
  \end{align}
and the elastic energy
  \begin{equation}
    F^E = \int_D  \frac{1}{2} K (\nabla_i Q_{jk}) \nabla_i Q_{jk} \; da,
  \end{equation}
which are both integrated over the material domain $D$.  Note that we have used
\begin{equation}
    \text{Tr}(\mathsf{Q}^2) = S^2/2.
    \label{r20}
\end{equation}  
We have also assumed isotropic elasticity with a single elastic constant.  We consider the more general case of anisotropic elasticity in Sect.~\ref{sec:anisotropic}.  The molecular tensor is then composed of a phase and elastic term
\begin{align}
  \mathsf{H} & = \mathsf{H}^P +  \mathsf{H}^E, \\
  \mathsf{H}^P  & =   -\mathsf{Q}(A + C\text{Tr}(\mathsf{Q}^2)) 
  =  -\mathsf{Q}\left(A + \frac{C}{2}S^2\right), \\ 
   \mathsf{H}^E & = \frac{1}{2}
 K \nabla^2 \textsf{Q}.
 \label{r22}
  \end{align}
 Note that the phase term  $\mathsf{H}^P$ does not violate nematic
 locking since it is proportional to $\mathsf{Q}$.  It is only the part of   $\mathsf{H}^E$ proportional to
 $\mathsf{U}$ that does.  We choose $C = -2A>0$ so that the integrand of Eq.~(\ref{r17}) is a double well, with $S = 1$ the location of the bottom of the right well.  Thus the phase free energy tends to drive the system toward alignment ($S = 1$).

 We now turn our attention to the fluid velocity $\mathbf{u}$, which
 is governed by the Navier-Stokes equation.  
\begin{equation}
 \frac{D}{Dt}\mathbf{u} =  \nabla_j \Pi_{ij},
  \label{r6}
\end{equation}
where we assume constant density $\rho = 1$, and where the stress tensor is
\begin{equation}
  \Pi = 2\eta \mathsf{E} - p \mathsf{I} +\Pi^E + \Pi^A.
\end{equation}
The first two terms are the viscous damping, with viscosity $\eta$,
and the pressure $p$.   The elastic stress is
\begin{align}
  \Pi^E & = -\mathsf{H} +
  [\mathsf{Q},\mathsf{H}] + 2 \text{Tr}(\mathsf{QH}) \mathsf{Q} -\nabla Q_{ij} \otimes \frac{\delta F}{\delta \nabla Q_{ij}},
\label{r4}
\end{align}
with the Ericksen stress term 
\begin{equation}
-\nabla Q_{ij} \otimes \frac{\delta F}{\delta \nabla Q_{ij}} = -K
\nabla Q_{ij} \otimes
\nabla Q_{ij}.
\end{equation}
Combining these results, the elastic stress becomes
\begin{align}
  \Pi^E & = K [-\nabla^2 \mathsf{Q} +
  [\mathsf{Q},\nabla^2 \mathsf{Q}] \nonumber \\
  & \; \; \; \;  + 2 \text{Tr}(\mathsf{Q} \nabla^2 \mathsf{Q}) \mathsf{Q}  -
\nabla Q_{ij} \otimes
\nabla Q_{ij}].
\label{r9}
\end{align}
Note that there is no stress associated with the phase free energy, as
 $\mathsf{H}^P$ contributes nothing to Eq.~(\ref{r4}).  
Finally, the active stress driving the motion is
\begin{equation}
\Pi^A = -\zeta \mathsf{Q},
\label{r5}
\end{equation}
with activity parameter $\zeta > 0$ for extensile dynamics. 

\begin{figure*}
\includegraphics[width = 2.0\columnwidth]{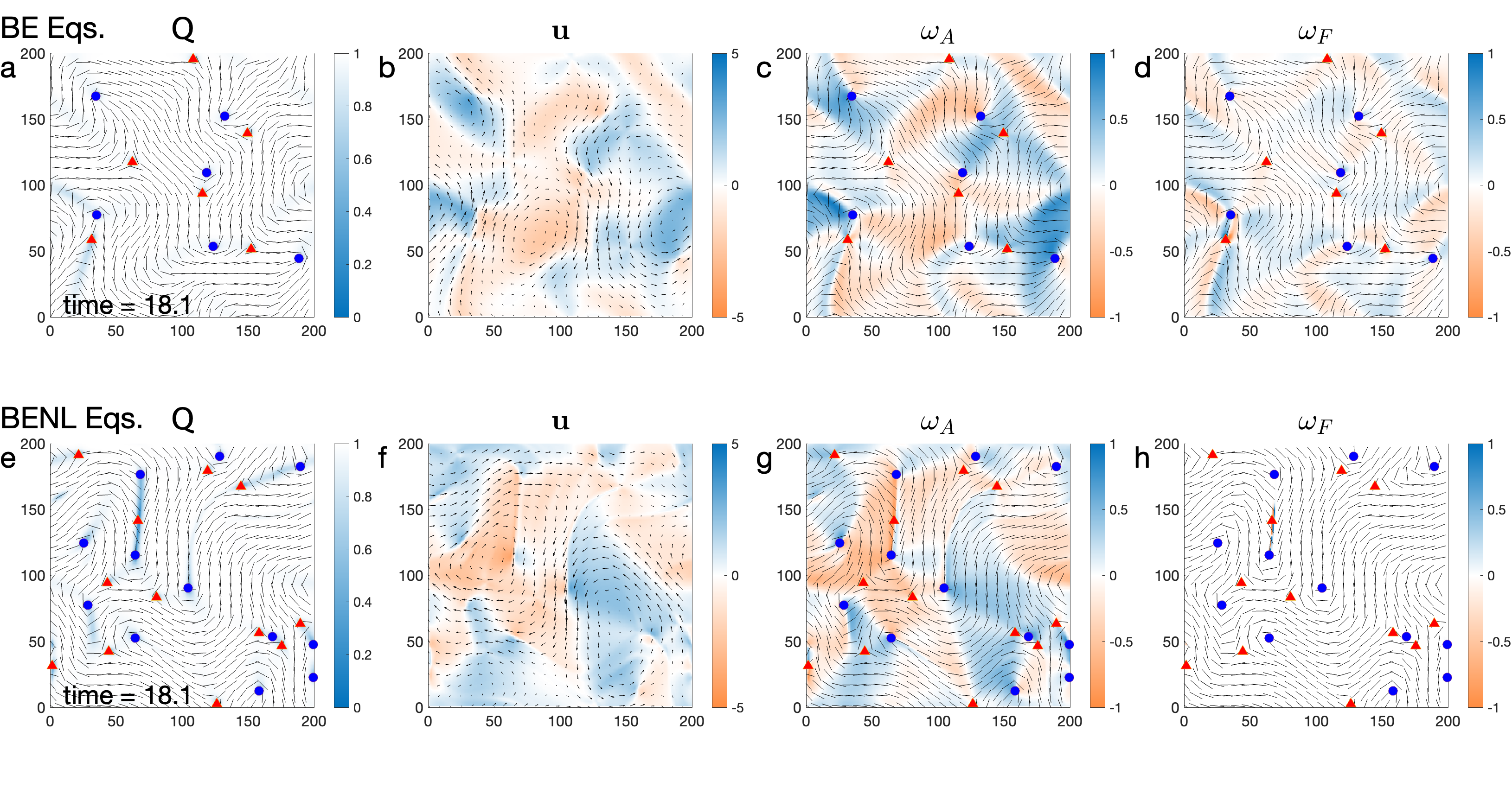}
\caption{\label{fig:BigFig_rand} Simulations of two model systems for active nematics.  Both simulations began with the same initial conditions, which consisted of randomly chosen directors.  Standard Beris-Edwards (BE) equations: a) Director field (lines) and $S$ (color). b) Velocity field with vorticity $\omega$ (color), c) $\omega_A$ (color), d) $\omega_F$  (color).   Plots e-h are the same as a-d except now using the Beris-Edwards equations with enhanced nematic locking (BENL). }     
\end{figure*}

\begin{figure}
 \centering
\includegraphics[width =  \columnwidth]{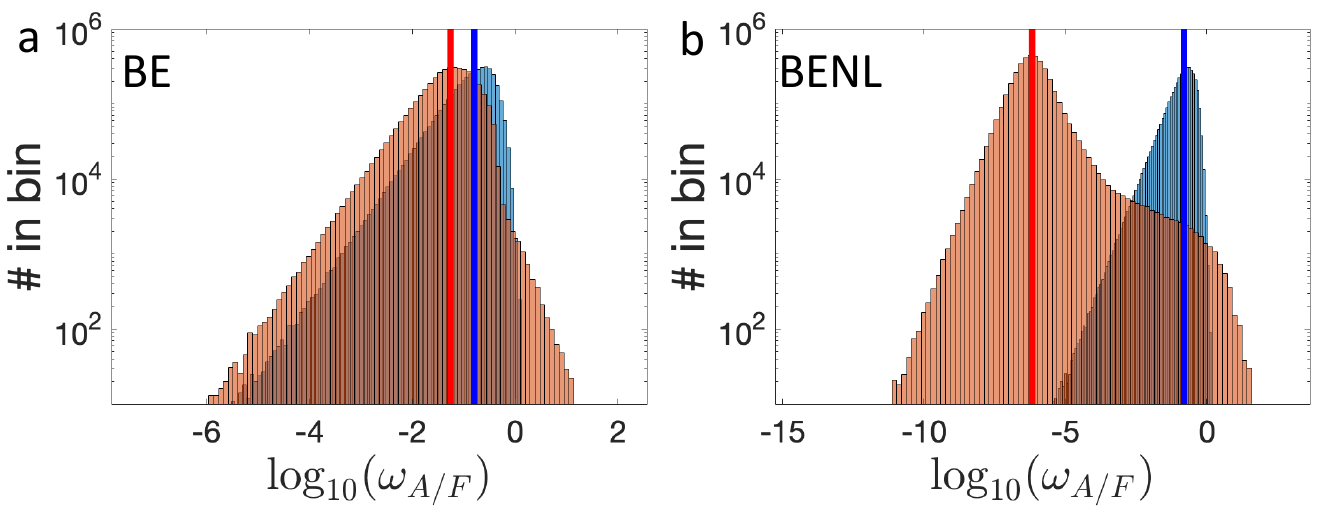}
\caption{ \label{fig:Histograms} Histograms of $|\omega_A|$ (blue) and $|\omega_F|$ (red) in log-space for 100 frames of simulation, computed from a) the standard Beris-Edwards equations and b) the Beris-Edwards equations with enhanced nematic locking.  The vertical lines are the medians of the distributions, where the color of the line matches the color of the distribution.}
\end{figure}

\begin{figure}
 \centering
\includegraphics[width =  \columnwidth]{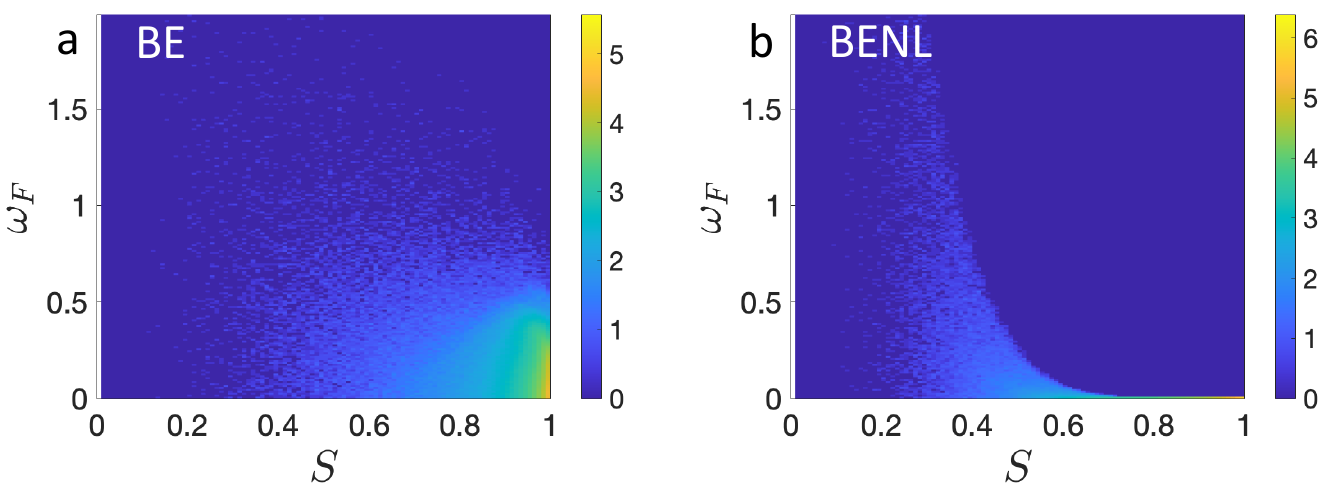}
\caption{ \label{fig:2DHistograms} The correlation between fracturing $\omega_F$ and the order parameter $S$ as represented by a histogram of $(S, \omega_F)$ values taken over all grid points in 100 frames of the simulation.  The colorbar is $\log_{10}$ of the number of occurrences of each $(S, \omega_F)$ value.  a) The standard Beris-Edwards equations.  b) The Beris-Edwards equations with enhanced nematic locking.}
\end{figure}

Figure~\ref{fig:BigFig_rand} shows a snapshot of the numerical solution of the Beris-Edwards model.  Figure~\ref{fig:BigFig_rand}a shows the director field and order parameter $S$, while Fig.~\ref{fig:BigFig_rand}b shows the velocity field.  As is well known, such simulations beautifully capture an essential feature of the experimental system, namely the creation and annihilation of topological defects.  See the corresponding supplemental video M1.  Parameters used for the simulation are: $\lambda = 1$, $\gamma = 5 \times 256$, $C = 256^2$, $K = 256^2$, $\eta = 2560$, and $\zeta = (256/3)^2$.  From these we can compute the following derived quantities.  The Reynold's number is $\text{Re} = K/\eta^2 = 0.01$ (i.e. we are in the Stokes limit), the active length is $\ell_a = \sqrt{K/\zeta} = 3$, and the nematic coherence length is $\ell_n = \sqrt{K/C} = 1$.  The integration domain is $200 \times 200$ with periodic boundary conditions.  We used an initial director field with randomly chosen orientations, with initial velocity $\mathbf{u} = 0$.   

\subsection{Does the Beris-Edwards model exhibit nematic locking?}

We saw that the $\textsf{H}^E$ term in the Beris-Edwards nematic transport equation has the potential to break nematic locking via fracturing of the material.  Here, we investigate to what degree it actually does.
To identify how prevalent fracturing is, we decompose the molecular tensor term $\mathsf{H}/\gamma$ into two parts
\begin{equation}
  \frac{1}{\gamma} \mathsf{H} = \frac{2}{\gamma S^2}[\text{Tr}(\mathsf{HQ}) \mathsf{Q} + \text{Tr}(\mathsf{HU}) \mathsf{U}],
  \label{r7}
 \end{equation}
which is possible because $\mathsf{H}$ is a $2 \times 2$ traceless, symmetric tensor and hence spanned by $\mathsf{Q}$ and $\mathsf{U}$. 
 The term proportional to $\mathsf{U}$ breaks nematic locking.  Its prefactor gives the fracturing angular velocity via Eq.~(\ref{r18})
\begin{equation}
   \omega_F = \frac{\text{Tr}(\mathsf{HU})}{\gamma S^2}
   = \frac{K \text{Tr}( \mathsf{U} \nabla^2 \mathsf{Q})}{2\gamma S^2}. 
\end{equation}
We compare $\omega_F$ to the angular velocity of the director due to passive advection $\omega_A$ [Eq.~(\ref{r3})].
Figures~\ref{fig:BigFig_rand}c and \ref{fig:BigFig_rand}d show $\omega_A(x,y)$ and $\omega_F(x,y)$.  The most striking aspect of this comparison is that $\omega_F(x,y)$ is very similar in spatial structure and magnitude to $\omega_A(x,y)$.  There are, however, two main differences.  First, $\omega_F$ tends to have the opposite sign to $\omega_A$, i.e. where $\omega_F$ is blue $\omega_A$ is red and vice versa.  Second, $\omega_F$ is somewhat smaller than $\omega_A$, though the same order of magnitude.  This can be seen quantitatively in the root-mean-square values, $\langle \omega_A(x,y) \rangle_\text{RMS} = 0.263$ and $\langle \omega_F(x,y) \rangle_\text{RMS} = 0.158$.   Thus, as the material is rotated via advection, it fractures to avoid the full effect of the rotation, i.e. fracturing decreasing the total amount of rotation.  The amount of fracturing rotation is roughly 60\% of the advective rotation.   For future reference, we also note the medians, $\text{median} (|\omega_A(x,y)|) = 0.1490$ and $\text{median}(|\omega_F(x,y)|) = 0.0687$.  Again, we see that the fracturing value is roughly half that of advection.  Finally, Fig.~\ref{fig:Histograms}a plots histograms of $|\omega_A|$ (blue) and $|\omega_F|$ (red) collected over 100 frames of the simulation (a duration of about 5.71, compared to the active time scale $t_a = K/(\zeta \eta) = 0.0035$).  The median values are shown as vertical lines.  The median of $|\omega_F|$ is roughly a third that of $|\omega_A|$.  We examine the correlation between $\omega_F$ and $S$ in the 2D histogram in Fig.~\ref{fig:2DHistograms}a.  This histogram is peaked at the bottom right, where we witness a considerable amount of fracturing ($\omega_F >0$) in the region where $S \approx 1$.  Physically, this corresponds to fracturing occurring in regions of high density.

\begin{figure}
 \centering
\includegraphics[width = \columnwidth]{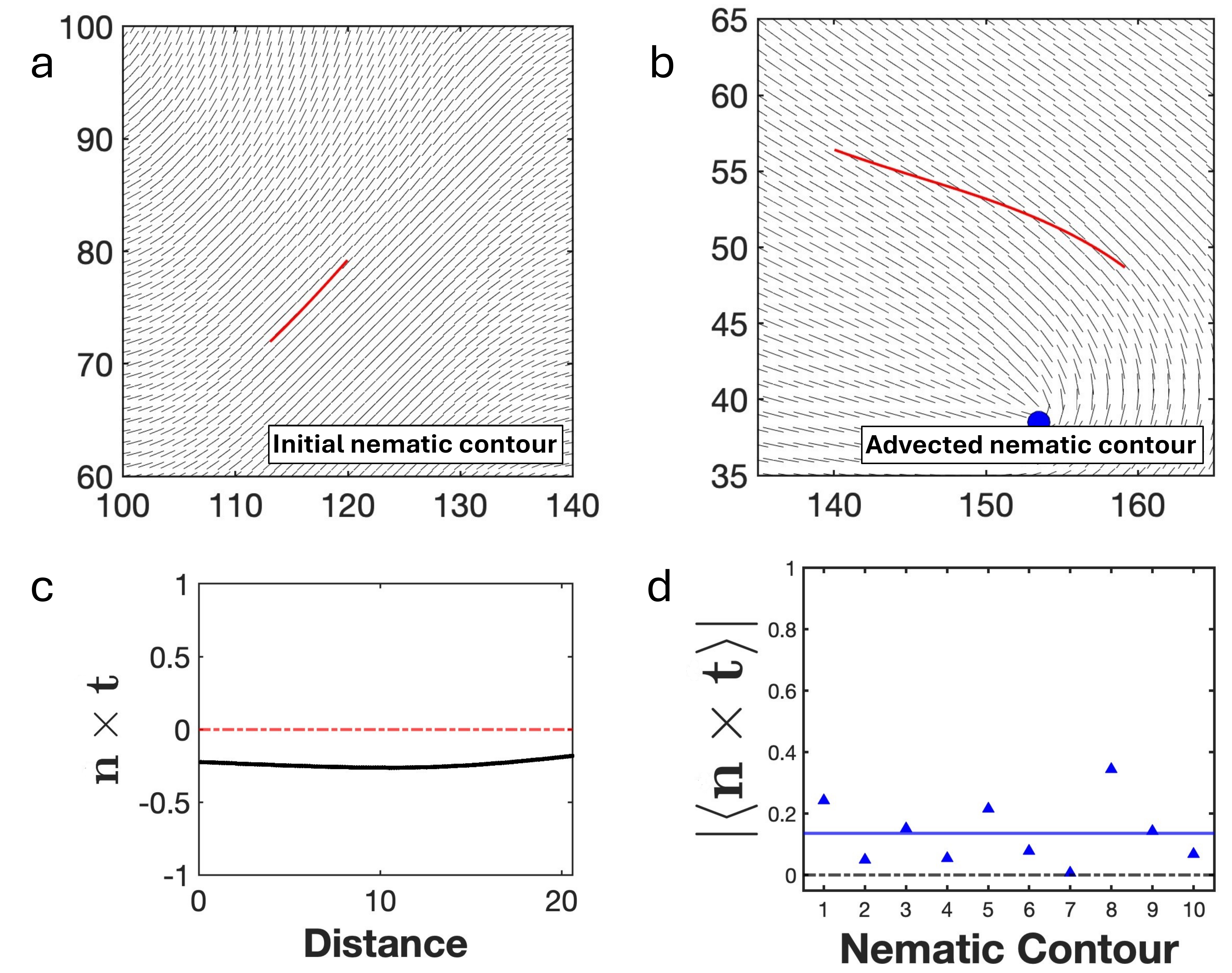}
\caption{ \label{fig:summaryTheory} Nematic contour analysis using the Beris-Edwards (BE) model. (a) An initial nematic contour is computed by spatially integrating the director field from a reference point at an initial time. (b) The passively advected nematic contour. (c) The cross product of the director field $\mathbf{n}$ and the curve tangent $\mathbf{t}$ along the advected curve. (d) Summary of nematic contour analysis using BE model for $10$ initialized nematic contours. The blue triangles are the absolute averages of $\mathbf{n} \times \mathbf{t}$ along the length of each advected curve. The blue line is the average of $|\langle \mathbf{n} \times \mathbf{t} \rangle|$ for 10 nematic contours. The cross product $\mathbf{n} \times \mathbf{t}$ shows a significant deviation from zero compared to what we observed in experiments.}
\end{figure}

Next, we measure the extent of fracturing in the Beris-Edwards model by passively advecting a nematic contour, as we did with the experimental data in Sect.~\ref{sec:Expt}.  Figure~\ref{fig:summaryTheory}a shows an initial nematic contour, and Fig.~\ref{fig:summaryTheory}b shows the corresponding passively advected curve once it has doubled in length. The cross product $\mathbf{n} \times \mathbf{t}$, computed along the advected curve, is also shown (Fig.~\ref{fig:summaryTheory}c).  It is clearly very far from zero, meaning that nematic locking has been substantially violated in this example.  In total 10 initial curves were advected.  (See supplemental Fig.~S2.)  Figure~\ref{fig:summaryTheory}d  shows $\mathbf{n} \times \mathbf{t}$ averaged over the length of each of the 10 advected curves. The average of all $|\langle \mathbf{n} \times \mathbf{t} \rangle|$ values is $0.135$ with a standard deviation of $0.106$. The average $|\langle \mathbf{n} \times \mathbf{t}\rangle|$ for the Beris-Edwards model is more than three times as large as $0.036 \pm 0.031$ observed in the experiments.  The comparison is even more stark when one considers that the experimental data suffers from inaccuracies in the director and velocity fields due to considerable error in extracting these fields.  

In summary, the two analyses in this section show that the traditional Beris-Edwards model exhibits considerable fracturing throughout the entire fluid domain, rather than localized at regions of high curvature.  To better model experimental observations of microtubule-based active nematics, in the next section, we shall modify the Beris-Edwards model to impose the nematic locking principle throughout the majority of the material, so that fracturing becomes more localized.

\section{The Beris-Edwards model with enhanced nematic locking (BENL)}

\label{sec:BENL}

\begin{figure} 
\includegraphics[width = .5\columnwidth]{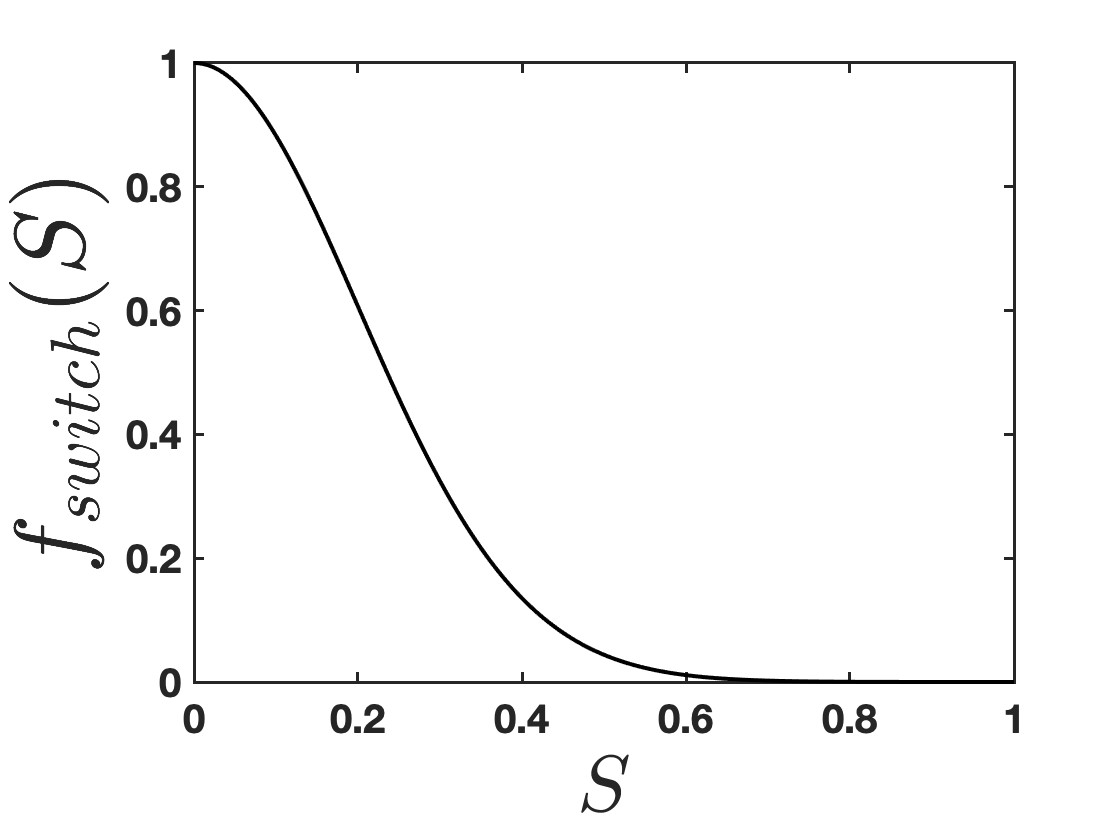}
\caption{\label{fig:switch}  The switch function for turning on fracturing when $S$ is close to zero. }
\end{figure}

The Beris-Edwards model violates nematic locking through the $\mathsf{U}$ component of the $\mathsf{H}$ tensor [Eq.~(\ref{r7})].  This component induces rotation of the director field due to fracturing.  Though this component is critical for defect creation and annihilation, it should be completely suppressed in most of the material where $S \approx 1$.  We thus modify the mobility, i.e. the prefactor, of the $\mathsf{U}$ term in $\mathsf{H}$ to depend on
 $S$ and only ``turn on'' when $S$ is reduced.  This is physically motivated by the fact that it is easier to fracture the material when it has lower density. 
 This is achieved with a highly nonlinear switch function, which ramps from (essentially) 0 at $S = 1$ to 1 at $S = 0$.  In principle, many different functional forms would produce the same qualitative behavior.  Here, we opt for
 a Gaussian switch function $f_{\text{switch}}(S) = \exp(-S^2/(2 \sigma^2))$, with $\sigma = 0.2$ (see Fig.~\ref{fig:switch}), which multiplies the $\mathsf{U}$ term, replacing $\mathsf{H}/\gamma$ in the
 nematic transport equation with
\begin{align}
 \frac{1}{\gamma} \mathsf{H}  & \rightarrow \frac{1}{\gamma} \frac{2}{S^2}[\text{Tr}(\mathsf{HQ}) \mathsf{Q} + e^{-S^2 /(2
     \sigma^2)} \text{Tr}(\mathsf{HU}) \mathsf{U}] \label{r14}\\ 
   &=
     \frac{1}{\gamma} \mathsf{H}+ \frac{1}{\gamma}\frac{2 (e^{-S^2 /(2
     \sigma^2)}-1)}{S^2} \text{Tr}(\mathsf{HU}) \mathsf{U}.
\label{r13}
\end{align}
The final expression is what is actually implemented in the simulation
code because it is a manifestly smooth function of position; note that $[\exp(-S^2/(2\sigma^2) - 1]/S^2$ is a smooth function
in $S$ despite the $S^2$ in the denominator, since the leading term in the Taylor series of the numerator is order $S^2$.  For completeness we now record the final modified nematic transport equation.
\begin{align}
  \frac{D}{D t}  \mathsf{Q}
 &= S \mathsf{E} +  [\mathsf{Q}, \Omega ] -
2 \text{Tr} (\mathsf{Q} \mathsf{E} ) \mathsf{Q}  + \frac{1}{\gamma} \mathsf{H} \nonumber \\
&+ \frac{1}{\gamma}\frac{2 (e^{-S^2 /(2
     \sigma^2)}-1)}{S^2} \text{Tr}(\mathsf{HU}) \mathsf{U}.
     \label{r15}
\end{align}
The only difference from the standard Beris-Edwards nematic transport equation is the final term (and the requirement that $\lambda = 1$).

Note that the fracture rotation rate, which is determined by the prefactor of the $\mathsf{U}$ term in Eq.~(\ref{r14}) via Eq.~(\ref{r17}), is now multiplied by $f_{\text{switch}}$, yielding
\begin{equation}
   \omega_F = e^{-S^2/(2 \sigma^2)} \frac{K \text{Tr}( \mathsf{U} \nabla^2 \mathsf{Q})}{2\gamma S^2}. 
\end{equation}

The modification in Eq.~(\ref{r14}) can be naturally understood in the context of gradient descent of the free energy (which can in turn be derived from the Onsager Variational Principle~\cite{Thiele21}).  In general, gradient descent gives the dynamics for a set of $m$ fields $X_\alpha$, $\alpha = 1...m$, according to
\begin{equation}
\frac{d}{dt} X_\alpha = -M_{\alpha \beta} \frac{\delta F}{\delta X_\beta},
\end{equation}
with a sum over $\beta$, and where $F$ is a free energy depending on the $X_\alpha$'s (and their derivatives) and $M_{\alpha \beta}$ is the mobility tensor.  This tensor must be symmetric and positive and may depend on the $X_\alpha$'s.  In our case, we have (ignoring the flow-coupling terms and working in the Lagrangian frame)
\begin{equation}
    \frac{D}{Dt}Q_{ij} = M_{(ij)(kl)} H_{kl},
\end{equation}
where we use a double index, i.e. $\alpha = (ij)$.  Then the standard Beris-Edwards choice for the mobility tensor is $M_{(ij)(kl)} = \delta_{ik} \delta_{jl}/\gamma$.  For enhanced nematic locking we use instead
\begin{equation}
  M_{(ij)(kl)} = \frac{2}{\gamma S^2}[Q_{ij}Q_{kl} + f_\text{switch}(S)U_{ij}U_{kl}].
\end{equation}
Since $S$ [Eq.~(\ref{r20})] and $\mathsf{U}$ [Eq.~(\ref{r19})] can be expressed in terms of $\mathsf{Q}$, $M_{(ij)(kl)}$ is a function of $\mathsf{Q}$.  It is also clearly symmetric and positive.  Thus, the modified Beris-Edwards equation should be viewed as replacing the rotational viscosity with a nonlinear, matrix-valued, $\mathsf{Q}$-dependent mobility.  Note that the molecular tensor $\mathsf{H}$ itself does not change, and hence the Navier-Stokes equation uses the same $\mathsf{H}$ as before.  

Figure~\ref{fig:BigFig_rand}e shows a frame of the simulation using Eq.~(\ref{r15}).  All parameters and initial conditions are identical to the standard Beris-Edwards simulation in Fig.~\ref{fig:BigFig_rand}a.  The overall qualitative structure of the defects, director field, and $S$ are the same.  Furthermore, defects are created and destroyed in a similar manner in the two simulations.  (See supplemental video M1.) 
 Similarly the velocity field in Fig.~\ref{fig:BigFig_rand}f is qualitatively similar to that in Fig.~\ref{fig:BigFig_rand}b, and the advective angular velocities in  Figs.~\ref{fig:BigFig_rand}g and \ref{fig:BigFig_rand}c are also similar.  Where there is a striking difference is in the fracturing angular velocity in Fig.~\ref{fig:BigFig_rand}h.  We see essentially zero fracturing throughout the medium except in the vicinity of a pair that is being created, where there is a streak of blue and red, indicating an abrupt fracture in the material.  In the video of the simulation, we also see locations where there is a fracture ahead of a $+1/2$ defect leading to a forward jump in the defect.  This is also frequently seen in experiments.

We now look at the RMS values: $\langle \omega_A(x,y) \rangle_{\text{RMS}} = 0.249$ and $\langle \omega_F(x,y) \rangle_{\text{RMS}}  = 0.605$.   Thus, the integrated rotation rate due to fracturing is actually much higher than the integrated advective rotation rate.  However, the medians are very different, $\text{median} (|\omega_A(x,y)|) = 0.168$ and $\text{median}(|\omega_F(x,y)|) = 7.14 \times 10^{-7}$.  Thus there is a large amount of fracturing localized in a very small region of the material.  This can also be seen in the histogram of $|\omega_F|$ (red) in Fig.~\ref{fig:Histograms}b.  There is a tail of $|\omega_F|$ that extends well to the right of the histogram of $|\omega_A|$.  The size of this tail, however, is several orders of magnitude smaller than the peak.  This tail corresponds to the localized regions where large amounts of fracturing take place.

We again examine the correlation between $\omega_F$ and $S$.  We note that the vertical fracturing streak in Fig.~\ref{fig:BigFig_rand}h aligns with a streak of blue in Fig.~\ref{fig:BigFig_rand}e.  Such correlations are seen multiple times in the supplemental video M1; fracturing streaks always coincide with streaks of suppressed values of $S$.  The correlation between fracturing and small values of $S$ is also seen in the 2D histogram in Fig.~\ref{fig:2DHistograms}b.  Compared to the BE simulation in Fig.~\ref{fig:2DHistograms}a, the fracturing ($\omega_F >0$) is pushed to lower values of $S$, physically representing lower densities.  In fact, there is a close relationship between the shape of the histogram at $S \approx 0.6$ and the switch function at $S \approx 0.6$ in Fig.~\ref{fig:switch}.  Fracturing could be pushed to even lower values of $S$ by decreasing the width $\sigma$ of the switch function.  

In addition to the differences in $\omega_F$, we also noticed a difference in the number of defects.  The standard Beris-Edwards model has on average about 6.4 $+1/2$ defects, whereas the BENL model has about 9.6 $+1/2$ defects on average, over the $200 \times 200$ domain.  The fluid speed and vorticity distributions appear nearly the same between the two simulations.

\begin{figure}
 \centering
\includegraphics[width =  \columnwidth]{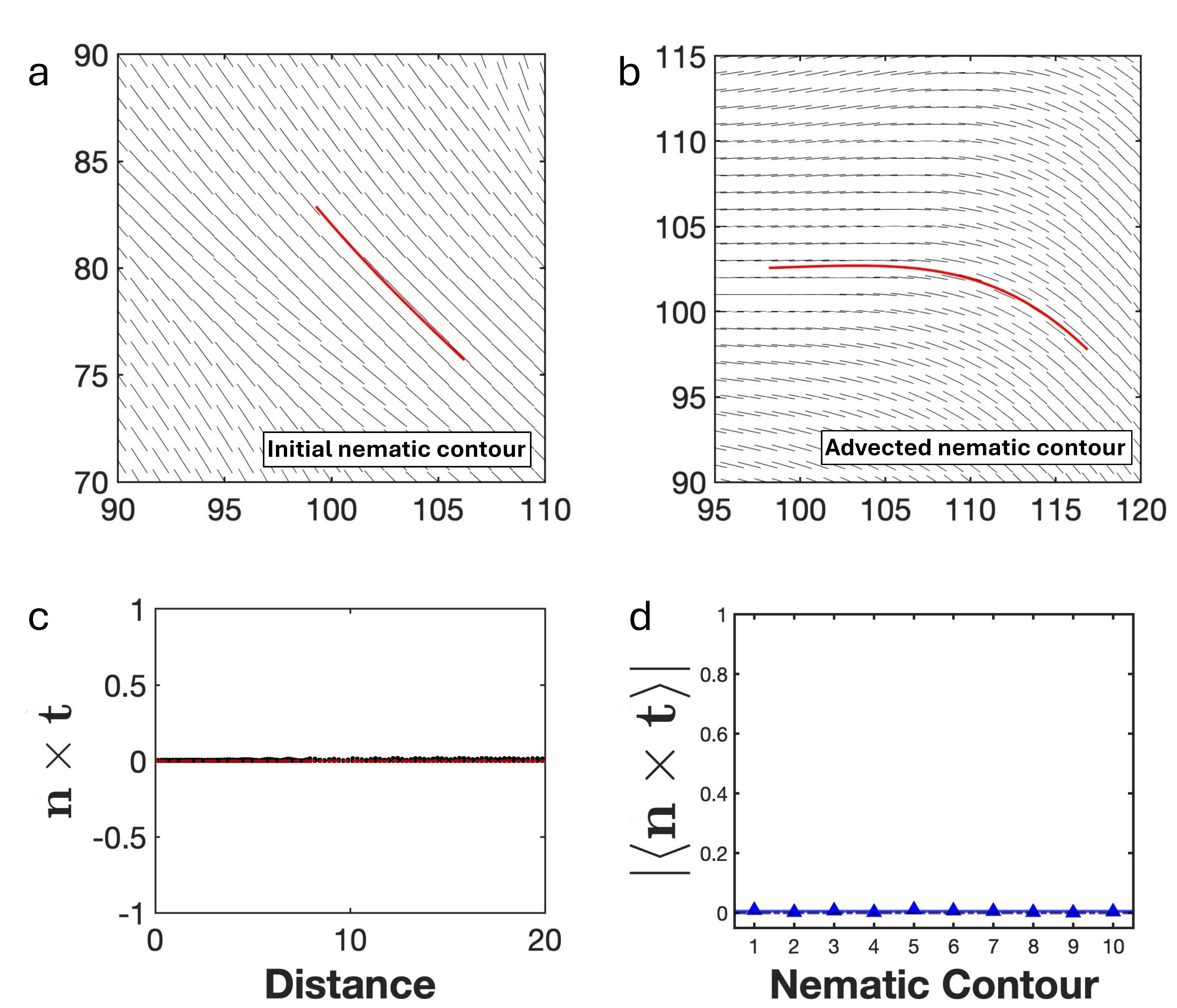}
\caption{ \label{fig:summaryBENL} Nematic contour analysis using the Beris-Edwards model with enhanced nematic locking (BENL). (a) An initial nematic contour is computed by spatially integrating the director field from a reference point. (b) The passively advected nematic contour. (c) The cross product of the director field $\mathbf{n}$ and the curve tangent $\mathbf{t}$ along the length of the advected curve. (d) Summary of nematic contour analysis using the BENL model for $10$ initial nematic contours. The blue triangles are the absolute averages of $\mathbf{n} \times \mathbf{t}$ along the length of each advected curve. The blue line is the average of $|\langle \mathbf{n} \times \mathbf{t} \rangle|$ for the 10 nematic contours. The cross product $\mathbf{n} \times \mathbf{t}$ is approximately zero, a trend that we observed in experiments.}
\end{figure}

\begin{figure}
\centering
  \includegraphics[width= 0.9 \columnwidth]{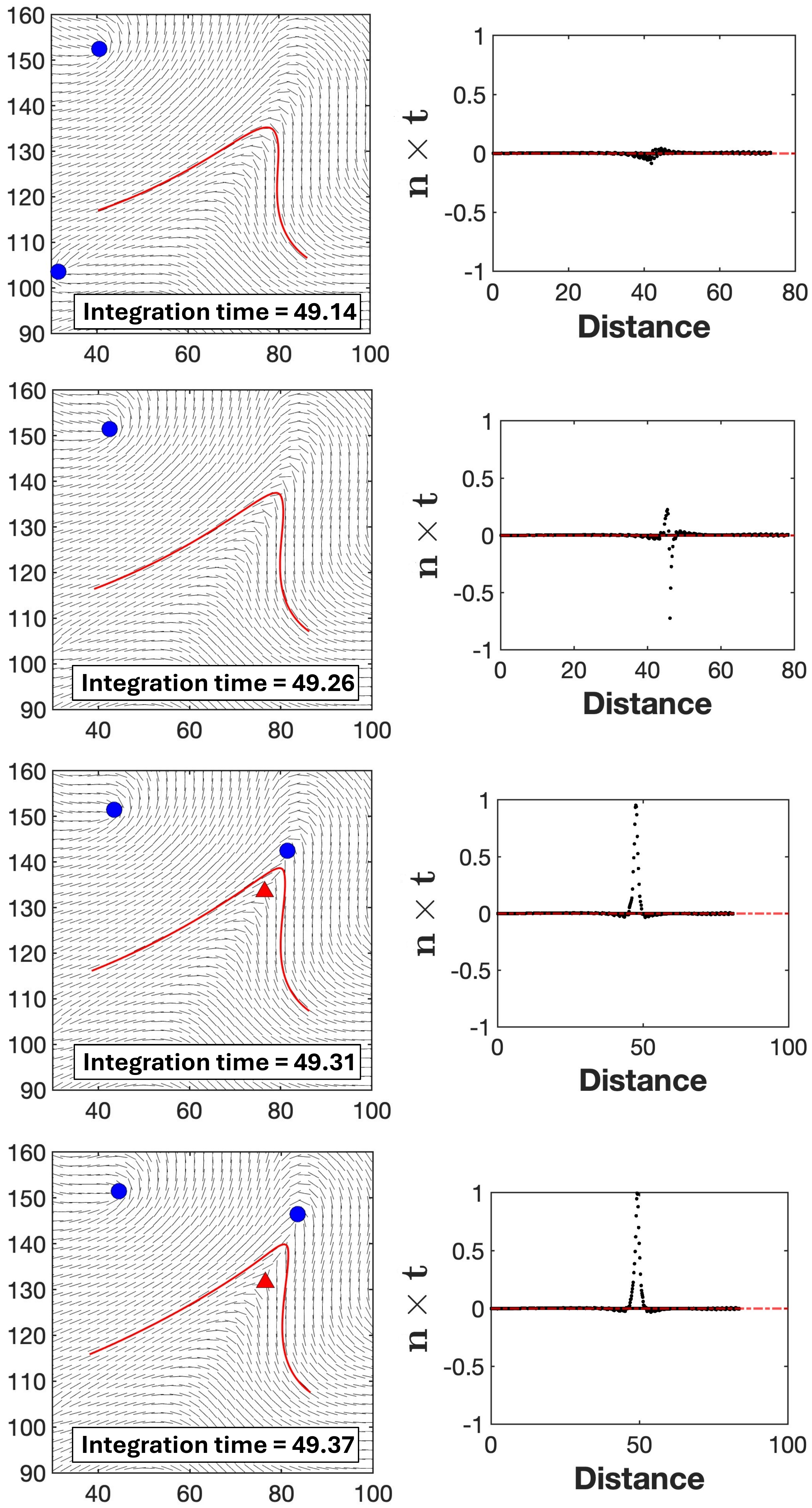}
 \centering
 \caption{ \label{fig:BENLfrac} Snapshots of the creation of a pair of topological defects using the BENL model. The red curve is a passively advected curve from an initial nematic contour that enters the fracture region where the defect pair is created.  In the plots of the right-hand column, the black dots are the values of $\mathbf{n} \times \mathbf{t}$ along the advected curve at four different time steps. The spike in $\mathbf{n} \times \mathbf{t}$ which develops at integration time $t = 49.31$ represents fracturing at the sharp bend of the advected curve.}
\end{figure}

We next apply the analysis of nematic contours to the modified Beris-Edwards model.  Figure~\ref{fig:summaryBENL}a shows an initial nematic contour and Fig.~\ref{fig:summaryBENL}b its passively advected counterpart, with double the initial length.  The cross product $\mathbf{n} \times \mathbf{t}$ computed along the advected curve is now essentially zero, meaning that the advected curve remains a nematic contour.  This process was again repeated for 10 initial curves.  Figure \ref{fig:summaryBENL} shows $|\langle \mathbf{n} \times \mathbf{t} \rangle|$ averaged over the length of each advected curve. The average of all $|\langle \mathbf{n} \times \mathbf{t} \rangle|$ values is $0.005$ with a standard deviation of $0.003$, which is approximately zero on a scale of zero to one. Next, we demonstrate that the fracture regions are localized in areas of high curvature by advecting a nematic contour through a region where a pair of topological defects is formed. Figure \ref{fig:BENLfrac} shows the snapshots of this evolution along with the plots of $\mathbf{n} \times \mathbf{t}$ at each time step. We observe a singularity develop in $\mathbf{n} \times \mathbf{t}$ at the sharp bend of the advected curve, similar to what we observed in the experimental images of Fig.~\ref{fig:FracSnaps}. It is also evident in Fig.~\ref{fig:BENLfrac} that  $\mathbf{n} \times \mathbf{t}$ settles down to zero away from the fracture region.

The BENL model can be viewed as giving rise to fracturing, including defect creation, in a three step process.   First the active dynamics creates regions of high curvature through chaotic stretching and folding.  Then, the energy cost in regions of high curvature drives down the value of $S$ locally via the first term in expression (\ref{r14}).  Finally, once $S$ is sufficiently small, local fracturing takes place  via the second term in expression (\ref{r14}).

One might argue that the functional dependence of the $\mathsf{U}$ prefactor in Eq.~(\ref{r14}) is unnecessary and that the same reduction in $\omega_F$ could be achieved by simply increasing the rotational viscosity $\gamma$, thereby suppressing the entire $\mathsf{H}$ term in the Beris-Edwards nematic transport equation.  In the limit of infinite $\gamma$, nematic locking is indeed recovered everywhere.  The problem lies in the singularities that would develop in the curvature of the nematic contours.  The very large $\gamma$ needed to adequately suppress the $\mathsf{H}$ term in the majority of the fluid would almost certainly lead to regions of unrealistically high curvature in the material before defect creation could relax this curvature.  The functional dependence of $f_{\text{switch}}$ on $S$ allows nematic locking in most of the material, but permits defect creation, and fracturing more generally, in regions where $S$ has been reduced due to high curvature.

\begin{figure*}
\includegraphics[width = 2.0\columnwidth]{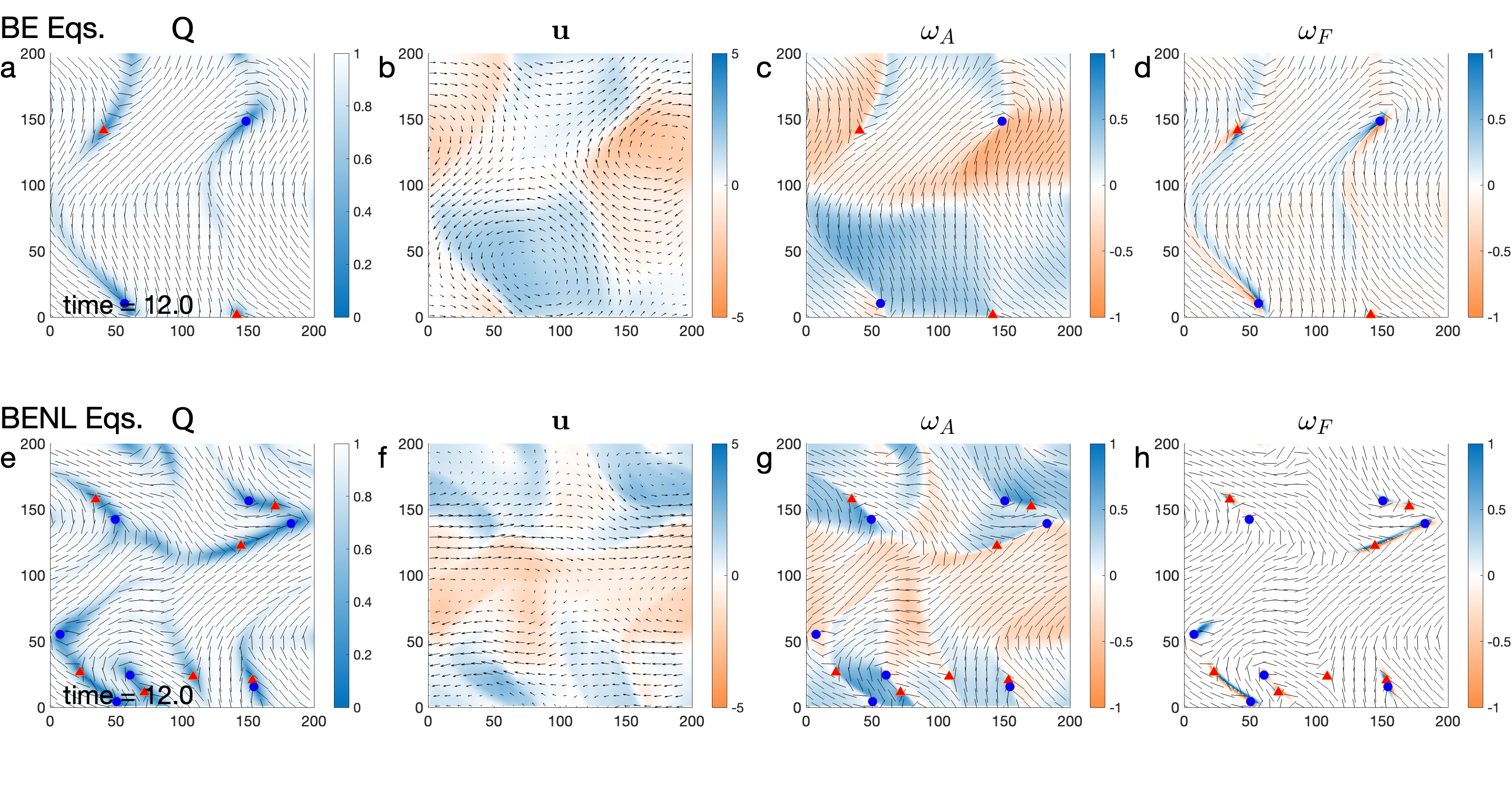}
\caption{\label{fig:BigFig_largecore_rand} Similar simulations as in Fig.~\ref{fig:BigFig_rand}, except that the value of $C$ has been decreased by a factor of 9, corresponding to an increase in the nematic coherence length $\ell_n$ by a factor of 3.}     
\end{figure*}

Finally, we show that the size of the fracturing domains can be increased by increasing the nematic coherence length $\ell_n$ from 1 to 3, which is achieved by decreasing $C$ from $256^2$ to $256^2/9$.  The results are shown in Fig.~\ref{fig:BigFig_largecore_rand} for both the standard Beris-Edwards model as well as the BENL model.  See also supplemental video M2. The core size is increased in both simulations.  The fracturing zones are now larger and easier to see in Fig.~\ref{fig:BigFig_largecore_rand}h as prominent streaks of blue next to red.  Such prominent streaks are also seen in Fig.~\ref{fig:BigFig_largecore_rand}d, but there is still a background haze of fracturing throughout the rest of the material as well.

\section{Comparison to the Beris-Edwards model with anisotropic elasticity}

\label{sec:anisotropic}

In equilibrium polymer nematic liquid crystals, the bend elastic constant $K_3$ is typically significantly larger than the splay elastic constant $K_1$ in the Frank elastic free energy \cite{mattoussi1989frank, zheng1984measurement, meyer1985measurements}. This is in contrast to the typical case in small-molecule nematic liquid crystals, where $K_3$ is usually the largest elastic constant \cite{DeGennes93}, as $K_1$ grows linearly with the chain length of semi-flexible polymers, whereas $K_3$ does not grow with chain length \cite{meyer1982polymer}.  
 Thus, one might suspect that we can suppress fracturing by introducing an elastic anisotropy.  We test this conjecture here using the standard Beris-Edwards model, but with anisotropic elasticity.

We use the following anisotropic elastic free energy 
  \begin{align}
    F^E &= \int_D  \frac{1}{2} \left[  L_1 (\nabla_i Q_{jk}) \nabla_i Q_{jk} 
    + L_2 Q_{ij} \nabla_i Q_{kl} \nabla_j Q_{kl} \right] \; da \\ 
     &= \int_D  \left[ K_3 S^2 |\nabla \times \mathbf{n}|^2 + 
     K_1 S^2 |\nabla \cdot \mathbf{n}|^2 \right] da, 
  \end{align}
where we have separated the free energy into bend and splay contributions with elastic constants $K_3$ and $K_1$ respectively. 
 Note in the final expression that we have ignored gradients in $S$ to focus on the director field.  The $L$ coefficients are expressed in terms of the $K$ coefficients as
 \begin{align}
     L_1 &= \frac{1}{2}(K_3 + K_1), \\
     L_2 &= \frac{1}{S}(K_3 - K_1).
 \end{align}
See the supplemental for a derivation.  Note that for isotropic elasticity $K_3 = K_1 = K$, so that $L_1 = K$ and $L_2 = 0$, and we recover the isotropic free energy Eq.~(\ref{r22}).  From Eq.~(\ref{r21}), we have
\begin{align}
    H_{ij}^E &= L_1 \nabla^2 Q_{ij} + L_2(\nabla_k Q_{kl} \nabla_l Q_{ij} - \frac{1}{2} \nabla_i Q_{kl} \nabla_j Q_{kl}    \nonumber \\
    & + \frac{1}{4} \nabla_m Q_{kl} \nabla_m Q_{kl} \delta_{ij}). 
    \label{r23}
\end{align}
Note that we can decompose $\mathsf{H}^E$ into bend and splay contributions $\mathsf{H}^B$ and $\mathsf{H}^S$.  The formula for $\mathsf{H}^B$ is obtained by substituting $L_1 \rightarrow K_3/2$ and $L_2 \rightarrow K_3/S$ in Eq.~(\ref{r23}).  Similarly,  $\mathsf{H}^S$ is obtained by substituting $L_1 \rightarrow K_1/2$ and $L_2 \rightarrow -K_1/S$.  We can then measure the amount of bend- and splay-induced fracturing according to $\omega_B = \text{Tr}(\mathsf{H}^B\mathsf{U})/(\gamma S^2)$ and $\omega_S = \text{Tr}(\mathsf{H}^S\mathsf{U})/(\gamma S^2)$.

\begin{figure*}
\includegraphics[width = 2.0\columnwidth]{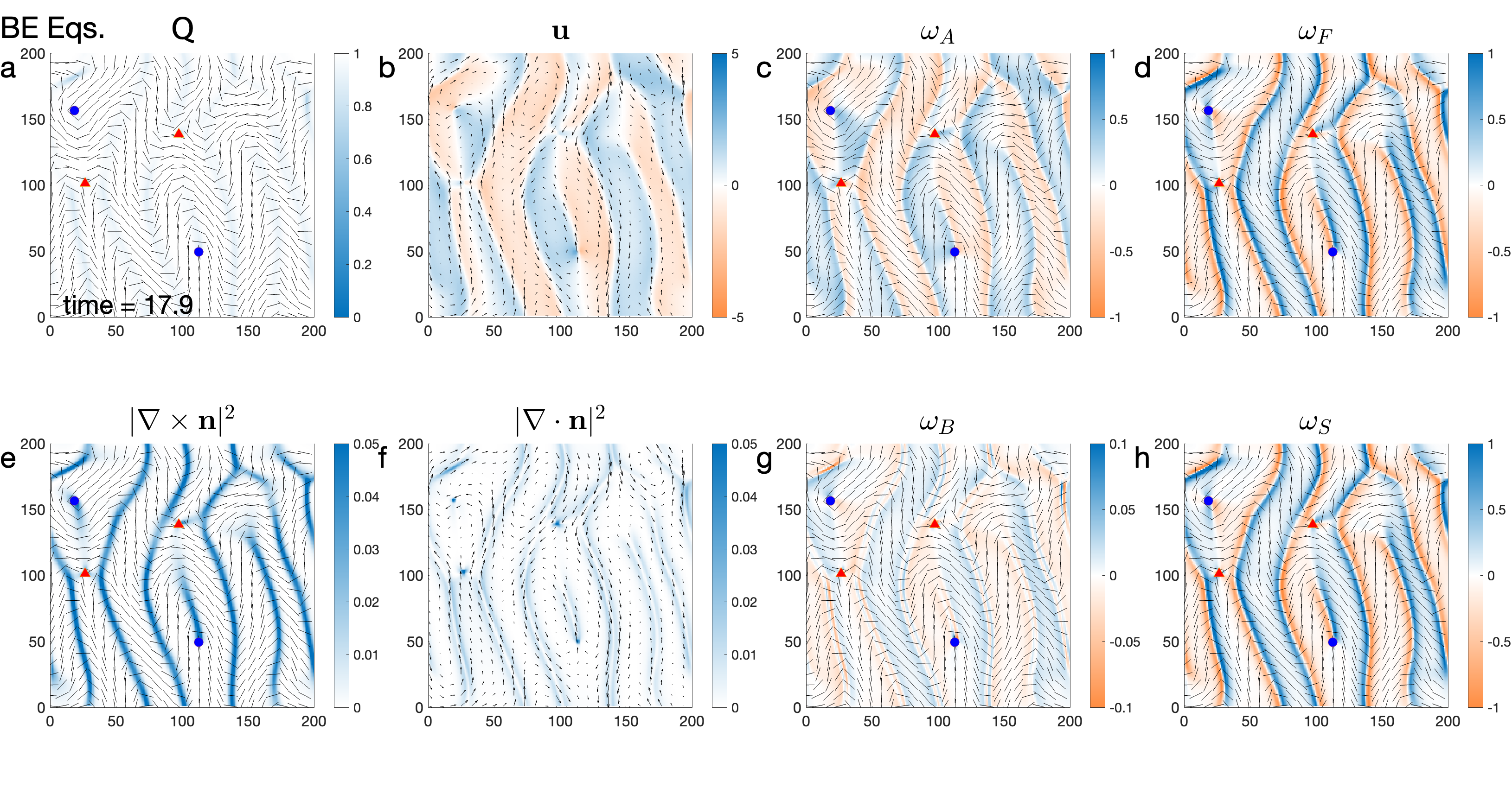}
\caption{\label{fig:BigFig_asymmetric}   a-d) Similar Beris-Edwards simulation as in Fig.~\ref{fig:BigFig_rand}a - \ref{fig:BigFig_rand}d, except that we use two distinct elastic constants: $K_3 = K_1/10$ and $K_1 = K = 256^2$.  e) The curl of $\mathbf{n}$ squared.  f) The divergence of $\mathbf{n}$ squared. 
 g) The fracturing rotation rate due to bend energy.  h)  The fracturing rotation rate due to splay energy.}     
\end{figure*}

With this anisotropic formalism, we ran simulations using the same initial conditions and parameters as Fig.~\ref{fig:BigFig_rand}a, except we use $K_3 = K_1/10$ and $K_1 = K = 256^2$.  Figure~\ref{fig:BigFig_asymmetric} shows the results.  The most notable fact is that $\omega_F$ (Fig.~\ref{fig:BigFig_asymmetric}d) is still substantial and distributed throughout the material.  We decompose $\omega_F$ into its constituents $\omega_B$ and $\omega_S$ in Figs.~\ref{fig:BigFig_asymmetric}g and \ref{fig:BigFig_asymmetric}h.  We do see that $\omega_B$ is suppressed relative to $\omega_A$ (note the smaller scale on the colorbar in Fig.~\ref{fig:BigFig_asymmetric}g), but it is still distributed throughout the material.  Thus, we see that anisotropic elasticity suffers the same deficiency as the standard isotropic Beris-Edwards model.

\section{stationary states}

\label{sec:fixed}

\begin{figure*}
\includegraphics[width = 2.0\columnwidth]{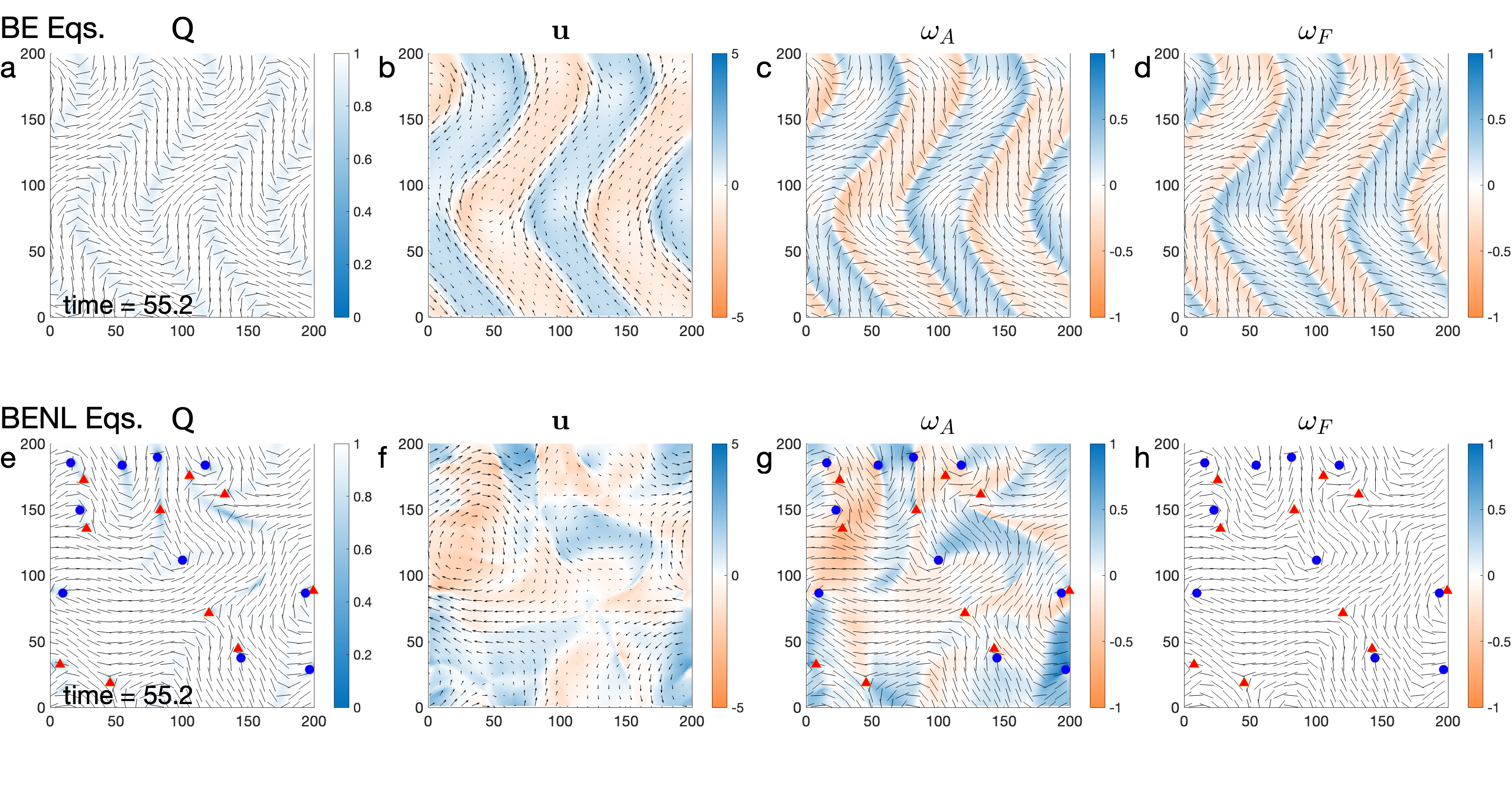}
\caption{\label{fig:BigFig_hor} Simulations of two model equations for active nematics.  Both simulations began with the same initial conditions, which consisted of a nearly horizontal director field.  Standard Beris-Edwards model, which has reached a stationary state that does not change: a) Director field (lines) and $S$ (color). b) Velocity field with vorticity $\omega$ (color), c) $\omega_A$ (color), d) $\omega_F$  (color).  The values in c and d are truncated at $\pm 1$.   Plots e-h are the same as a-d except now using the Beris-Edwards model with enhanced nematic locking. }     
\end{figure*}


\begin{figure}
 \centering
\includegraphics[width =  \columnwidth]{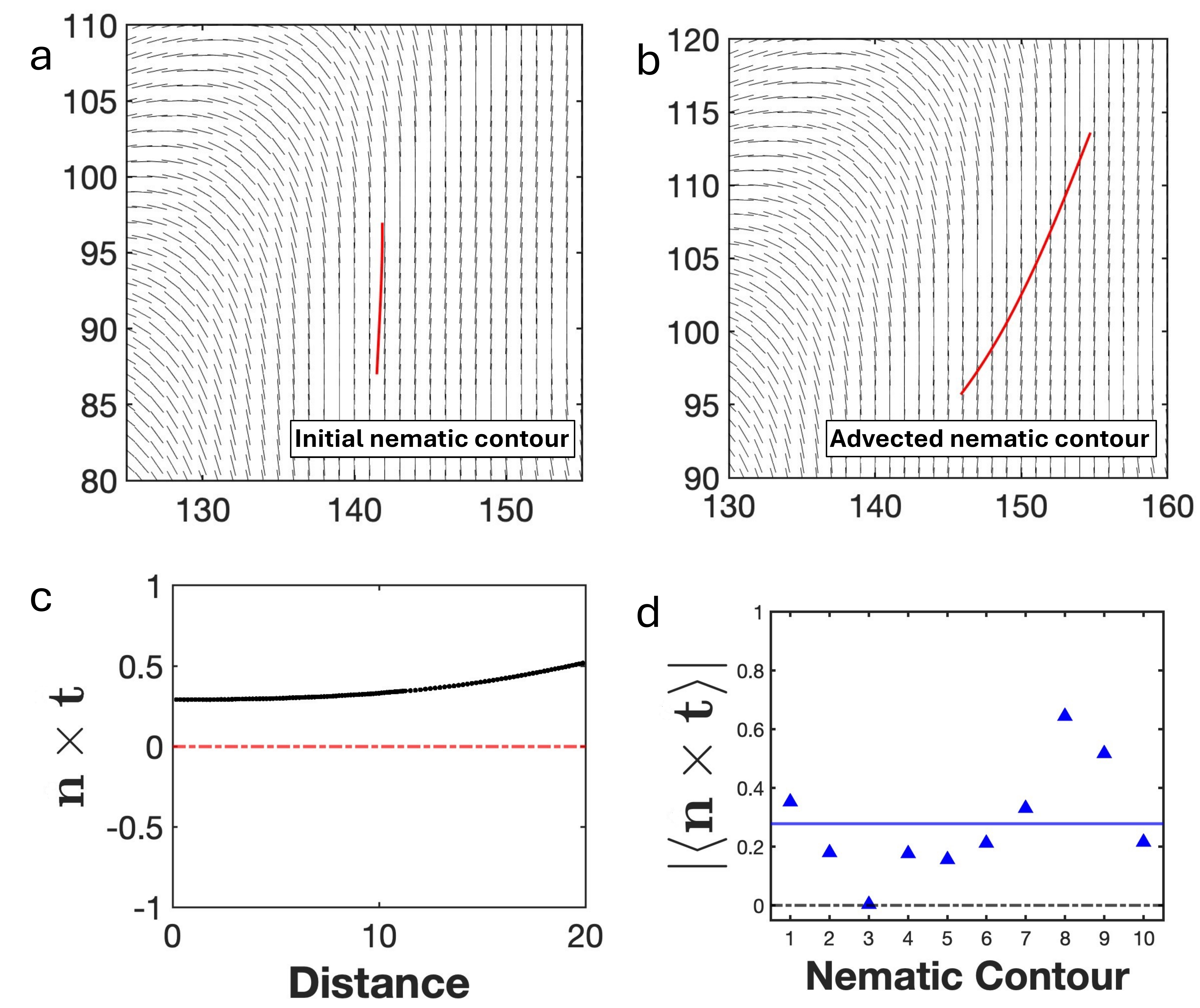}
\caption{ \label{fig:summaryBEFS} Nematic contour analysis using a stationary state that occurs in simulations of standard BE model. (a) An initial nematic contour is computed by spatially integrating the director field at an initial time. (b) The passively advected nematic contour. (c) The cross product of the director field, $\mathbf{n}$ and the curve tangent, $\mathbf{t}$ along the length of the advected contour shown in (b). (d) Summary of nematic contour analysis for $10$ initialized nematic contours. The blue triangles are the absolute averages of $\mathbf{n} \times \mathbf{t}$ along the length of each advected curve. The blue line is the average of $|\langle \mathbf{n} \times \mathbf{t} \rangle|$ for 10 nematic contours, and the black breaking line represents $|\langle \mathbf{n} \times \mathbf{t} \rangle| = 0$. The cross product, $\mathbf{n} \times \mathbf{t}$, shows a massive deviation from zero.}
\end{figure}

We now consider an extreme example of fracturing that can occur in simulations and show that the BENL model eliminates such states.  
Some simulations of the standard Beris-Edwards model exhibit non-turbulent solutions with no defects and with stationary flows and stationary nematic structure, i.e. both the velocity and director fields do not change in time~\cite{Shendruk17,Wagner22,Mitchell24}.  These states tend to occur as the activity is lowered.  However, we know of no examples of such states seen in actual experiments with microtubule-based active nematics.  We provide here an explanation for why such states may not occur in experiments---stationary flows with stationary director fields violate the nematic locking principle.
The argument proceeds as follows.  In a bounded domain, a stationary, area-preserving flow generates closed streamlines, in which passive tracers move along periodic orbits.  A passively advected needle-shaped nematogen may either align tangent or transverse to its own periodic orbit.  In the tangential case, this would imply that the  nematic contour would be the same as the streamline.  The nematic contour would thus form an invariant closed loop.  This violates the extensile nature of the dynamics, which would require that such a loop expand in time.  In the transverse case, in order for the needle to return to its original orientation after one period, the periods of adjacent streamlines must be the same.  This would mean that all streamlines in a given region would have the same period, which in turn implies that any interval of a nematic contour would return to itself after one such period.  But this also violates the extensile nature of the nematic.  

Note that the above argument breaks down if the active contribution to the Navier-Stokes equation vanishes, i.e. if $\nabla_j \Pi_{ij}^A = 0$.  Assuming $S = 1$, this condition can be expressed as
\begin{align}
0 &= \nabla_j \Pi_{ij}^A /\zeta
= \nabla_j (n_i n_j) 
= n_j \nabla_j n_i + (\nabla \cdot \mathbf{n})n_i \nonumber \\
&= n_j (\nabla_j n_i - \nabla_i n_j)  + (\nabla \cdot \mathbf{n})n_i \nonumber\\
&= n_j \epsilon_{ji} (\nabla \times \mathbf{n})  + (\nabla \cdot \mathbf{n})n_i \nonumber\\
&=  (\nabla \times \mathbf{n}) n^\perp_i + (\nabla \cdot \mathbf{n})n_i,
\end{align}
where $\epsilon_{ji}$ is the antisymmetric $2\times 2$ tensor with $\epsilon_{xy} = 1$.  Thus, the active contribution vanishes if and only if both $\nabla \times \mathbf{n}$ and $\nabla \cdot \mathbf{n}$ vanish.  On a square with periodic boundary conditions, this means $\mathbf{n}$ must be constant.  We conclude that an active nematic that respects the nematic locking principle can not exhibit stationary states, except the trivial case of constant $\mathbf{n}$ and zero velocity.

We now demonstrate the above result numerically.  We first integrate the standard Beris-Edwards equations using the same parameters as the simulation in Fig.~\ref{fig:BigFig_rand}a, but using a different initial director field; we now use an initial horizontal director field, with a small random perturbation.   Instead of exhibiting active turbulence, the director field converges to a stationary zig-zag state with no defects, shown in Fig.~\ref{fig:BigFig_hor}a, with a stationary velocity field, shown in Fig.~\ref{fig:BigFig_hor}b, forming channels that match the zig-zag pattern and that alternate between flowing up and down.  See also supplemental material video M3.  Note that the system exhibits significant advective torque (Fig.~\ref{fig:BigFig_hor}c) that is essentially entirely canceled by fracturing torque (Fig.~\ref{fig:BigFig_hor}d).  Quantitatively, we have 
$\langle \omega_A(x,y) \rangle_\text{RMS} = 0.190$ and $\langle \omega_F(x,y) \rangle_\text{RMS} = 0.179$, and   
$\text{median} (|\omega_A(x,y)|) = 0.133$ and $\text{median}(|\omega_F(x,y)|) = 0.139$.  Thus the amount of fracturing rotation is essentially the same as the amount of advective rotation.

We now integrate the BENL equations using the same parameters and initial conditions as the standard Beris-Edwards equations.   The BENL equations quickly go unstable and produce active turbulence, essentially identical to what is seen in Fig.~\ref{fig:BigFig_rand}e.  Thus, the BENL equations fail to converge to the stationary zig-zag pattern. Lastly, we apply nematic contour advection tests to the stationary states of the BE equations (see Fig.~\ref{fig:summaryBEFS}). The value of $|\langle \mathbf{n} \times \mathbf{t} \rangle|$ for the stationary state is $0.279$ with a standard deviation of $0.189$. The large deviation from a nematic contour can be easily observed in Fig.~\ref{fig:summaryBEFS}b as the final advected curve is transverse to the nematic director field. 

\begin{figure*}
\includegraphics[width = 2.0\columnwidth]{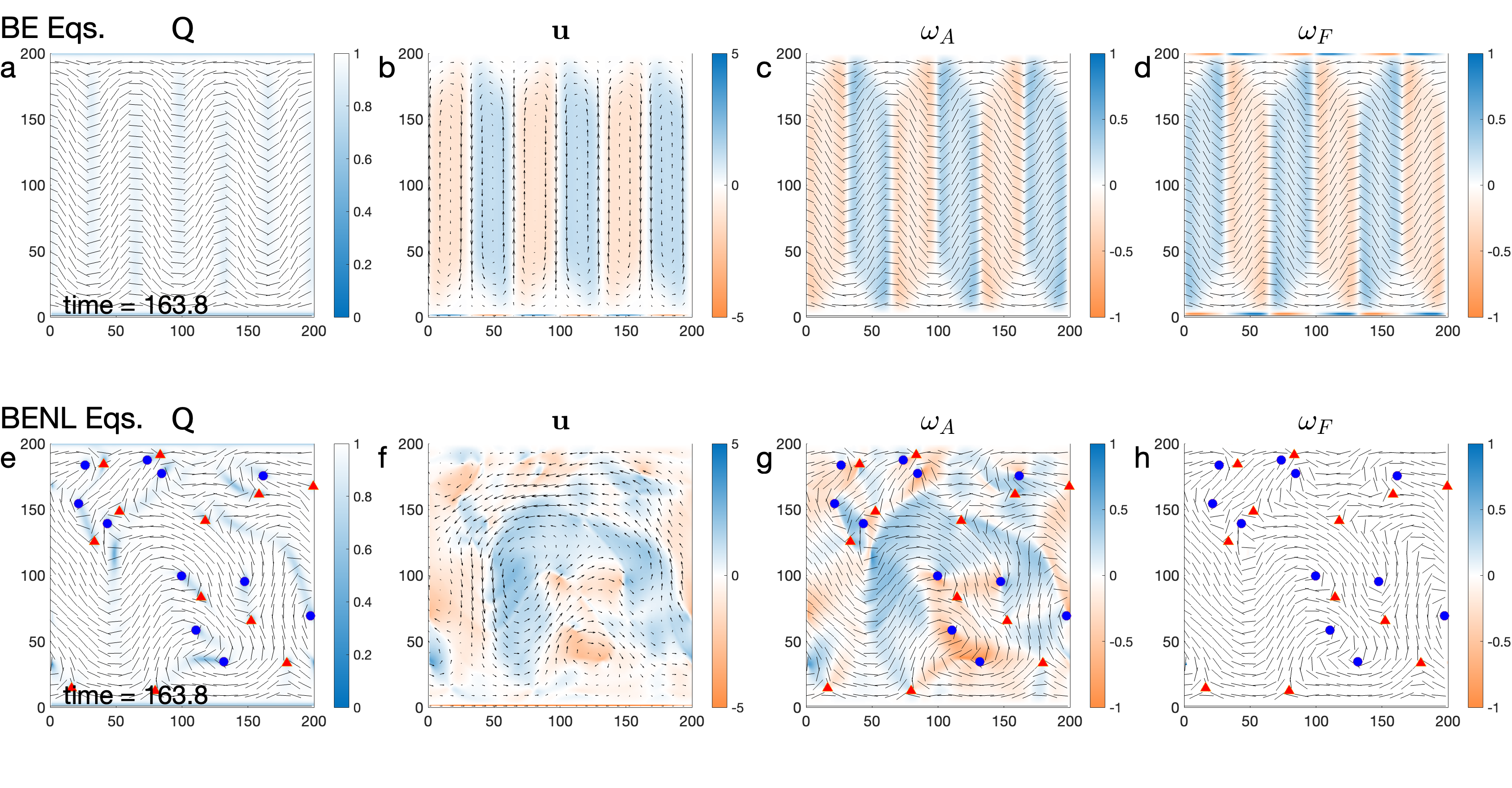}
\caption{\label{fig:BigFig_channel} Simulations of two model equations for active nematics similar to Fig.~\ref{fig:BigFig_hor}, except now in a channel geometry.  Both simulations began with the same initial conditions, which consisted of a nearly horizontal director field.  Standard Beris-Edwards model, which has reached a stationary state that does not change: a) Director field (lines) and $S$ (color). b) Velocity field with vorticity $\omega$ (color), c) $\omega_A$ (color), d) $\omega_F$  (color).  The values in c and d are truncated at $\pm 1$.   Plots e-h are the same as a-d except now using the Beris-Edwards model with enhanced nematic locking. }     
\end{figure*}

As a second example of a stationary state, we consider a channel geometry, with a hard wall on the top and bottom and periodic boundary conditions on the left and right.  This system was shown to produce stationary states in the simulations of Ref.~\onlinecite{Shendruk17}, along with the now well known dancing disclinations.  We confirm in Figs.~\ref{fig:BigFig_channel}a-d that our Beris-Edwards simulations produce such a stationary state, using exactly the same parameters as Fig.~\ref{fig:BigFig_hor}.  The director field forms a sinusoid-like pattern, while the fluid velocity forms a series of alternating vorticies.  Examining $\omega^A$ and $\omega^F$, they again look nearly identical but with their signs reversed.  Thus, the stationary state solution again violates nematic locking to a large degree.  The BENL simulation in Figs.~\ref{fig:BigFig_channel}e-h, however, exhibits active turbulence and preserves nematic locking throughout the majority of the fluid, as seen by the predominantly white background in Fig.~\ref{fig:BigFig_channel}h.

\section{Conclusions}

We have shown here how a simple, perhaps even obvious, principle---that a nematic contour advected forward remains a nematic contour---can lead to a reformulation of the standard Beris-Edwards equations.  The key takeaway is that the elastic energy contribution to the nematic transport equation can be decomposed into two terms, one which preserves nematic locking and one which violates it.  The term that violates nematic locking must therefore be suppressed throughout the majority of the material.  Doing so leads to simulations that restrict fracturing of the microtubules to localized strips of high curvatures (and suppressed $S$), consistent with experimental observation.

Note that we do not expect our switch function approach to apply universally to all active nematic systems, e.g. bacterial swarms, where the director can more freely rotate.  However, it would be interesting to study whether the nematic locking approach applies to other physical systems, such as the actin-myosin system~\cite{Kumar18}.

\acknowledgments

We would like to thank Linda Hirst and Mattia Serra for extensive discussions of this work.  We would also like to thank Linnea Lemma and Zvonimir Dogic for providing access to their experimental data in Ref.~\onlinecite{Serra23}.  This work was financially supported by the US Department of Energy under grant DE-SC0025803, by the US National Science Foundation through the Center of Research Excellence in Science and Technology: Center for Cellular and Biomolecular Machines at the University of California Merced (HRD-1547848 and HRD-2112675) and through NSF grant DMR-2225543, and finally by the University of California Office of the President under grant M25PL8991 (the UC Active Matter Hub). 

\bibliography{MyBibDeskBib}

\end{document}